\documentclass[11pt]{article}

\usepackage[margin=1in]{geometry}
\usepackage{lmodern}        % scalable fonts, so large elements (e.g. the draft watermark) render at full size
\usepackage{amsmath}
\usepackage{amssymb}
\usepackage{booktabs}
\usepackage{graphicx}
\graphicspath{{Fig/}}

\usepackage[colorlinks=true, citecolor=blue, linkcolor=blue, urlcolor=blue]{hyperref}
\usepackage{caption}        % needed for the SM \captionsetup below
\usepackage{float}          % needed for the SM's [H] figure/table placement

\usepackage{amsthm}
\theoremstyle{plain}

\theoremstyle{definition}

\theoremstyle{remark}

\usepackage{draftwatermark}
\SetWatermarkText{Draft}
\SetWatermarkScale{1}   % lmodern now scales properly, so this no longer needs to be inflated

\title{AI in Search Reduces Publisher Referrals Without Improving User Experience: Experimental Evidence}

\author{
Stephanie~T.~Wang$^{1,\dagger,\ast}$,
Jeffrey~Gleason$^{2,\dagger}$,
Yakov~Bart$^{3}$,
Christo~Wilson$^{2}$,
Danaé~Metaxa$^{1}$
\\[1em]
$^{1}$Computer and Information Science, University of Pennsylvania, Pennsylvania, United States\\
$^{2}$Khoury College of Computer Sciences, Northeastern University, Massachusetts, United States\\
$^{3}$D'Amore-McKim School of Business, Northeastern University, Massachusetts, United States
\\[1em]
$^{\dagger}$These authors contributed equally to this work.\\
$^{\ast}$Corresponding author: \href{mailto:stephtw@engineering.upenn.edu}{stephtw@engineering.upenn.edu}
}
\date{}

\begin{document}

\maketitle

\begin{abstract}
The integration of generative AI into web search delivers synthesized answers to user queries, changing how people navigate and assess information, while raising concerns about the downstream impacts on publishers who supply the underlying content. We conduct a preregistered field experiment (N=1,100) on Google Search, the dominant online search platform, to estimate the causal effects of AI Overviews and AI Mode on user behavior, perceptions, and publisher traffic. We show that removing AI Overviews and AI Mode increases click-through rates to publishers, while an AI Mode-only experience reduces click-through rates and erodes user experience and trust in information found on Google. These findings show that integrating generative AI into web search reshapes online attention, with economic consequences for the online publishers that sustain both search platforms and the overall information ecosystem.
\end{abstract}


\bigskip
\noindent\textbf{Significance statement.} Generative AI is fundamentally changing search and how billions of people access information online. In a preregistered randomized controlled trial with participants using Google Search in their everyday browsing, we find that removing AI features in Google Search increases clicks to third-party publishers, while Google's conversational AI search reduces clicks and worsens user experience. Our findings show how integrating generative AI into search can reshape how users engage with the broader web, providing more information directly in search while reducing engagement with external sources, with implications for the economic sustainability of online publishers.

\bigskip

\section{Introduction}
% Changes to Google SERP
Google Search has come a long way since its early days as a list of ``ten blue links.'' In May 2024, Google introduced AI Overviews (AIO), which use large language models (LLMs) to synthesize sources and generate summary answers to search queries directly on the results page~\cite{reid_generative_2024}. One year later, Google launched AI Mode, a conversational search experience powered by Gemini that generates direct AI responses to queries and supports multi-turn, follow-up questions~\cite{stein_expanding_2025}.  

% Publisher and Google tensions
The design of the search engine results page (SERP) matters because it shapes how users assess the credibility of information sources~\cite{lurie_investigating_2018} and distributes clicks and attention between third-party publishers and Google's own properties~\cite{gleason_google_2023, pape_is_2026}. The introduction of AI into search has raised these stakes and increasingly strained the relationship between Google and third-party publishers: publishers view AIO and AI Mode as \textit{substitutes} for their content and report falling referral traffic from Google Search~\cite{simonetti_news_2025, bearne_publishers_2025, chapekis2025google, conger_google_2026}. Google disputes these claims, arguing that AI features are \textit{complements} that have kept overall referral traffic steady, improved click quality, and expanded opportunities for websites to be surfaced through longer, more complex queries~\cite{reid_ai_2025}. On the other hand, users may benefit from the convenience of having their questions answered directly on the search results page in what is known as ``good abandonment''~\cite{diriye_leaving_2012, li_good_2009}, creating a three-way tension in which user convenience, publisher sustainability, and Google's consolidation of informational authority do not easily align.  

This dynamic of AI substituting for third-party content is not unique to search. Beyond search specifically, generative AI summaries have been shown to substitute for engagement with particular third-party publishers: ChatGPT's release coincided with significant declines in Stack Overflow visits and questions, particularly for topics ChatGPT excels at and with no corresponding drop in post quality, suggesting genuine displacement rather than a culling of low-quality content~\cite{burtch_consequences_2024, del_rio-chanona_large_2024}, while Wikipedia articles overlapping with ChatGPT's content saw similar declines in editing and viewership~\cite{lyu_wikipedia_2025}.  

Observational studies about the effect of LLM adoption on search activity and engagement with online sources more broadly have found mixed effects. Some find substitution, where LLM use replaces traditional web search, leading to a net decrease in search activity and downstream web traffic~\cite{padilla_impact_2025}. Others find complementarity, where LLM use augments rather than displaces traditional search, increasing the number of websites visited and occurring alongside rather than instead of traditional search~\cite{gholami2026beyond}. Evidence on AI features embedded directly in search remains comparatively scarce: one study finds that AIOs reduced English Wikipedia traffic by 15\%~\cite{khosravi2026impact}, with findings consistent with substitution being strongest when short answers satisfy users' informational intent. AI Mode, Google's fully conversational search experience, remains understudied despite representing the most significant departure yet from traditional Google Search design.  

% news publishers and platforms 
% This debate follows extensive prior literature on platforms and news publishers. ~\cite{chiou_content_2017, Athey_2021_TheIO}~\cite{luca_does_2015}

A related body of work examines the properties of generative search outputs themselves and their effects on users. Generative search outputs answer queries with fluency and a sense of credibility, yet reduce transparency and sourcing ability~\cite{memon_search_2024, shah_envisioning_2024}. These systems produce consequential errors in high-stakes domains~\cite{hu_auditing_2026, williams_why_2024}, exhibit systematic source-selection biases~\cite{chen_generative_2025, yang_news_2025, huang2026answer}, and reduce hedging language, making responses appear more informative and confident than their sources warrant~\cite{huang2026answer, liu_evaluating_2023, narayanan_venkit_search_2025}.
These properties of generative search have motivated studies examining downstream effects on user trust~\cite{li_human_2025}, belief formation~\cite{xu_ai_2025, sharma_generative_2024}, and general satisfaction~\cite{spatharioti_effects_2025, xu_chatgpt_2023}. An RCT found that users trust GenAI search answers less than traditional search; notably, citations increased trust even when hallucinated, while uncertainty highlighting reduced it~\cite{li_human_2025}. This tracks with a separate finding that users prefer responses with more citations regardless of their credibility~\cite{miroyan_search_2026}. The incongruence between model confidence and source credibility has motivated experiments on uncertainty-highlighting interventions, which reduce over-reliance on incorrect responses~\cite{kim_im_2024, kim_fostering_2025, spatharioti_effects_2025}. AI summaries have been shown to affect attitudes and behavioral intentions~\cite{xu_ai_2025}, and AI search can increase selective exposure and reinforce pre-existing beliefs~\cite{sharma_generative_2024}. LLM-based search has also been found to increase overall satisfaction with the search experience~\cite{spatharioti_effects_2025, xu_chatgpt_2023}. 

\subsection*{Research Questions and Hypotheses}

Existing work documents meaningful consequences of generative AI in search for users, publishers, and platforms, but relies primarily on observational data and controlled laboratory experiments. How, and to what extent, generative AI in search shapes user behavior, perceptions, and publisher traffic in naturalistic, real-world settings remains an open question. In particular, it is unclear whether AI in search primarily facilitates users’ discovery of third-party publishers or increasingly substitutes for it. To address this gap, we conduct a preregistered field experiment~\cite{osf_anonymous} in which participants are randomly assigned to one of three conditions via a browser extension~\cite{wang_lower_2024, lam_sociotechnical_2023, piccardi2025reranking} that manipulates the availability of AI in search features~\cite{google_ai_2026} and captures behavioral and attitudinal outcomes: 1) \textit{No AI Search}, which hides AI Overviews and AI Mode; 2) \textit{Current Search}, which makes no changes; and 3) \textit{AI Mode Search}, which redirects all searches to AI Mode.

We preregistered three primary hypotheses (H1--H3) and four secondary hypotheses (H4--H7), each predicting an effect of AI Mode Search and No AI Search relative to Current Search.
Our first three hypotheses focus on outcomes with direct relevance to users, publishers, and Google:

\paragraph{Hypothesis 1.} AI Mode Search will decrease overall trust in responses relative to Current Search, while No AI Search will increase it. This prediction builds on prior work documenting reduced trust in AI-generated search responses~\cite{li_human_2025}.

\paragraph{Hypothesis 2.} AI Mode Search will decrease click-through rate to external sites relative to Current Search, while No AI Search will increase it. This prediction is motivated by publisher reports of declining referral traffic~\cite{simonetti_news_2025} and observational data on AI Overviews~\cite{chapekis2025google}.

\paragraph{Hypothesis 3.} AI Mode Search will increase the number of search sessions relative to Current Search, while No AI Search will decrease it. This prediction is based on Google's own public claims that AI features increase search engagement~\cite{reid_ai_2025}.

Beyond these primary outcomes, we examine four secondary hypotheses related to user experience and search behavior: 

\paragraph{Hypothesis 4.} AI Mode Search will increase perceptions of responses as personalized and relevant relative to Current Search, while No AI Search will decrease these perceptions, motivated by Google's own claims about AI-driven personalization~\cite{google_personal_intelligence_search}. In the context of news-related search, AI Mode Search will decrease preference for and trust in AI Mode relative to Current Search, while No AI Search will have no impact on these outcomes. We did not have directional priors for how AI Mode Search and No AI Search affect satisfaction with, usefulness of, and agency over search responses, relative to Current Search.

\paragraph{Hypothesis 5.} AI Mode Search will decrease total outbound clicks, clicks to news sites, Wikipedia~\cite{khosravi2026impact} and Reddit, and ads relative to Current Search, while No AI Search will increase these outcomes.

\paragraph{Hypothesis 6.} AI Mode Search will increase minutes per session and the fraction of question-form searches relative to Current Search, while No AI Search will decrease them. We did not have a directional prior for how AI Mode Search and No AI Search affect the number of searches per session, relative to Current Search.

\paragraph{Hypothesis 7.} AI Mode Search will increase substitution to other search engines (Bing, DuckDuckGo, Yahoo) relative to Current Search, while No AI Search will have no impact on substitution.
\section{Experimental Methods and Data}
\label{methods}
We recruited participants from two channels: research recruitment platform Prolific, and Northeastern students enrolled in a work-study program with one of the authors. Participants were recruited in waves between March 17 and March 19, 2026. We limited recruitment to US-based participants, and screened for individuals who (1) were 18 years or older, (2) used Google Chrome as their primary browser, and (3) used Google Search as their primary search engine. After giving informed consent, participants completed a pre-survey (Section~\ref{si:pre-survey}) that elicited their familiarity with and sentiment towards LLM applications, as well as perceived trust, usefulness, satisfaction, agency, and personalization/relevance of information-seeking on Google. Participants then installed our browser extension and experienced three days of the baseline, Current Search condition. This period allowed us to measure pre-treatment versions of behavioral outcomes like search sessions per day and click-through rate. 

After three days, participants were randomly assigned to one of our three search conditions: (1) No AI Search, (2) Current Search, or (3) AI Mode Search. The No AI condition hid Google's AI Overviews, including those at the top of the results page, in the middle of the results page, and nested within People Also Ask components. AI Mode searches were also redirected to general search. The Current Search condition made no modifications to Google Search: AI Overviews may be present and AI Mode is accessible. The AI Mode condition redirected all searches to AI Mode. Participants experienced their assigned conditions for seven days. At the end of the seven-day period, we sent participants a post-survey that asked about perceptions of information-seeking on Google (Section~\ref{si:post-survey}) and presented head-to-head comparisons of AI Mode and Current Search responses in the context of news queries (Figure~\ref{fig:head_to_head_news}).

Our preregistration specified a target sample size of 1,200 participants, based on power analyses conducted using data from a pilot experiment~\cite{osf_anonymous}. To reach this target, a total of 1,444 participants (1,387 from Prolific and 57 from Northeastern) were enrolled, each completing the pre-survey and installing our browser extension. Of these, $N=1,100$ made at least one search during the 7-day experiment period and were invited to take the post-experiment survey. A total of $N=956$ participants completed the post-experiment survey, 343 of whom were in the No AI Search condition, 304 in Current Search, and 309 in AI Mode Search. Treatment groups and the control group were balanced across covariates, and there was no evidence of differential attrition across the three conditions (Section~\ref{si:validity}). Following our pre-registration, treatment effects on behavioral outcomes are estimated among participants eligible for the post-experiment survey ($N=1,100$). Treatment effects on survey outcomes are estimated among participants who completed the survey ($N=956$).

Prolific participants received \$10 upon completing the study and Northeastern work-study students were paid \$17 per hour. The analysis sample (i.e., participants who qualified for the post-survey) skewed younger (75\% under 45), highly educated (87\% with at least some college), and left-leaning (58\% Democrat vs. 20\% Republican). Race and ethnicity and gender distributions were similar to the US Census~\cite{census_us_2025}. 50\% of participants identified as women, 48\% as men, and 2\% as another gender. Participants were predominantly white (68\%), followed by Black or African American (14\%), Hispanic or Latino (11\%), and Southeast or East Asian (11\%). Race and ethnicity was multiple-choice. During the baseline period, participants averaged 9.2 searches per day (sd = 14.3), 4.0 search sessions per day (sd = 4.7), and 4.1 clicks per day (sd = 7.0). The SI Appendix (Section~\ref{si:robustness}) describes how we grouped searches into sessions, and validates that results are robust to different definitions. We observed that 36\% of searches triggered AI Overviews (Figure \ref{fig:ai_overview_visible_time}) and participants used AI Mode for 0.6\% of their searches (Figure \ref{fig:ai_mode_compliance_search}).
\section{Results}
\label{results}
This section presents our findings about the impacts of AI search features on user behaviors and user perceptions. For each outcome variable, we first describe the impact of AI Mode Search and then the impact of No AI Search.

Each result is labeled with its corresponding pre-registered~\cite{osf_anonymous} hypothesis. The section ends with a qualitative exploration of open-ended feedback from participants in our AI Mode condition.

\subsection*{Compliance and Attrition}
\label{results:compliance-attrition}
During the experiment period, 94.7\% of searches in the AI Mode group were successfully routed to AI Mode. In the No AI condition, 90\% of AI Overviews were successfully hidden on the first intervention day, but the success rate declined to 0\% as the experiment progressed (Figure \ref{fig:no_ai_compliance_time}). The cause of this breakdown was a change to AI Overview HTML rolled out by Google during the study period that broke our extension's logic for identifying and hiding AI Overviews. Overall, 51.1\% of AI Overviews were successfully hidden and the median participant in this condition had 50\% of AI Overviews hidden (Figure \ref{fig:no_ai_compliance_users}). Our pre-registration specified that we would report local average treatment effect (LATE) estimates in the case of non-compliance. In the No AI Search condition, we measure compliance as the fraction of AI Overviews that were successfully hidden for each participant. In the AI Mode Search condition, we measure compliance as the fraction of searches that were routed to AI Mode.  

A related concern for causal interpretation is differential attrition across conditions: if assignment to AI Mode caused higher participant attrition than other conditions, our AI Mode estimates would reflect a selected subsample rather than the full randomized population, and would be better treated as descriptive than causal. We do not find evidence of this in our study: survey non-completion (36.6\% vs.\ 32.3 -- 32.5\% in the other conditions, $p=0.292$) and zero-search attrition (26.5\% vs.\ 20.7 -- 24.0\%, $p=0.109$) were both directionally higher but not statistically significant in the AI Mode group (Section~\ref{si:attrition}).

Throughout the rest of this section, we reference intent-to-treat (ITT) effects when discussing AI Mode Search (i.e., ``assignment to AI Mode") and local average treatment effects (LATEs) when discussing No AI Search (``exposure to No AI"). The corresponding figures and tables show both estimates. Each result is labeled with its corresponding pre-registered hypothesis.

\subsection*{Effects on User Behaviors}
\begin{figure*}[ht]
	\centering
    \includegraphics[width=0.75\textwidth]{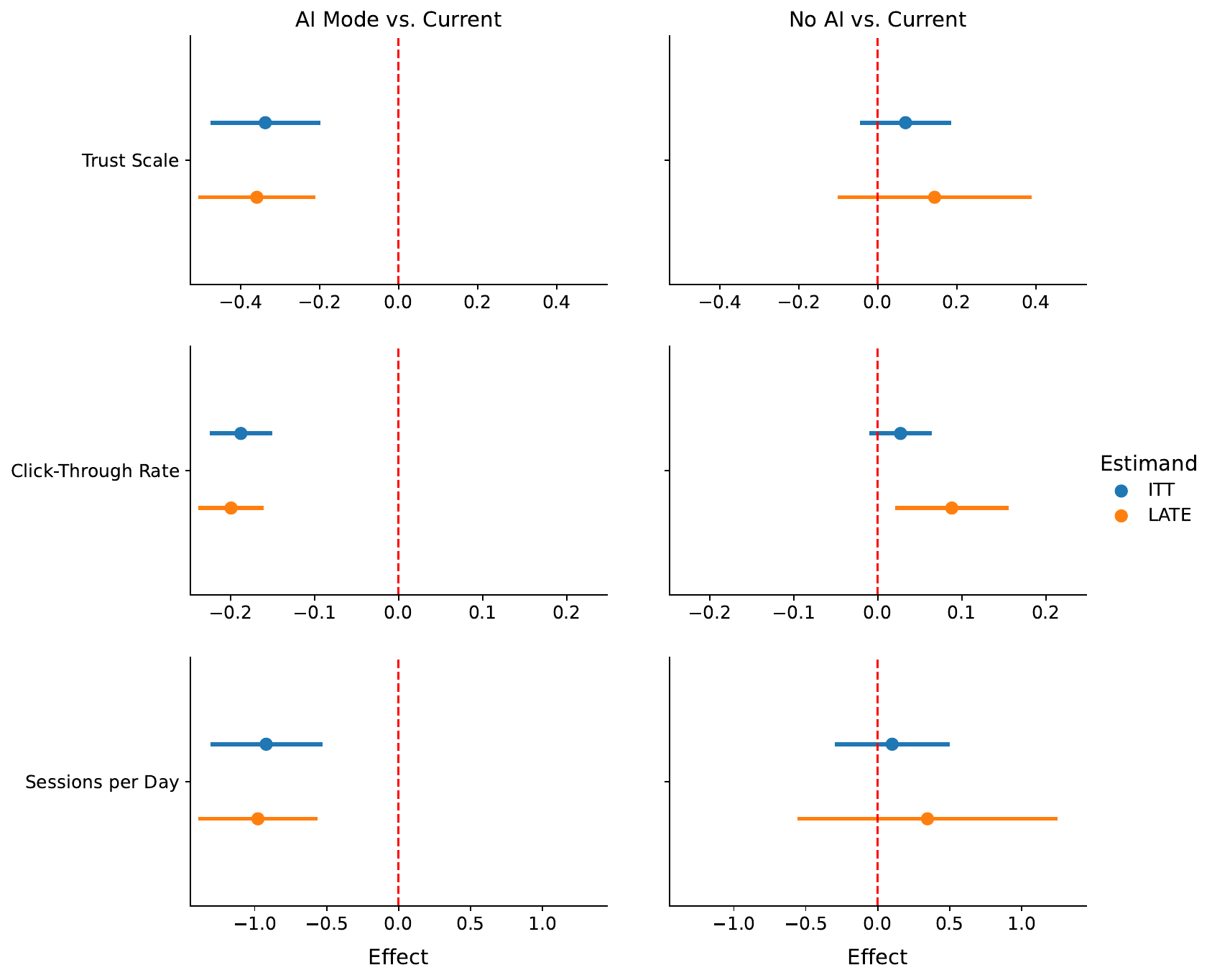} 
    \caption{\textbf{Effects of AI Mode Search and No AI Search on Primary Outcomes.}}
    \label{fig:primary}
\end{figure*}
\paragraph{Click-Through Rate (H2).} Assignment to AI Mode Search significantly reduced click-through rate by -18.8 percentage points (pp) [95\% CI: -22.2pp, -15.3pp; $p< 0.001$], consistent with H2a (Figure \ref{fig:primary} and Table \ref{tab:primary}). Conversely, exposure to No AI Search increased click-through rate by 8.8pp [95\% CI: 2.3pp, 15.3pp; $p = 0.008$], consistent with H2b. Our pre-registered robustness check analyzing click-through rate at the search-level also shows a positive impact of No AI Search (Figure \ref{fig:search_level}, top panel).

\paragraph{Search Sessions per Day (H3)} Assignment to AI Mode Search significantly reduced search sessions per day by -0.92 sessions [95\% CI: -1.30, -0.55; $p< 0.001$], contrary to H3a, which predicted an increase. We do not find evidence that exposure to No AI Search impacted search sessions per day [95\% CI: -0.55, 1.24; $p = 0.175$], failing to support H3b.

\begin{figure}[ht]
	\centering
    \includegraphics[width=\textwidth]{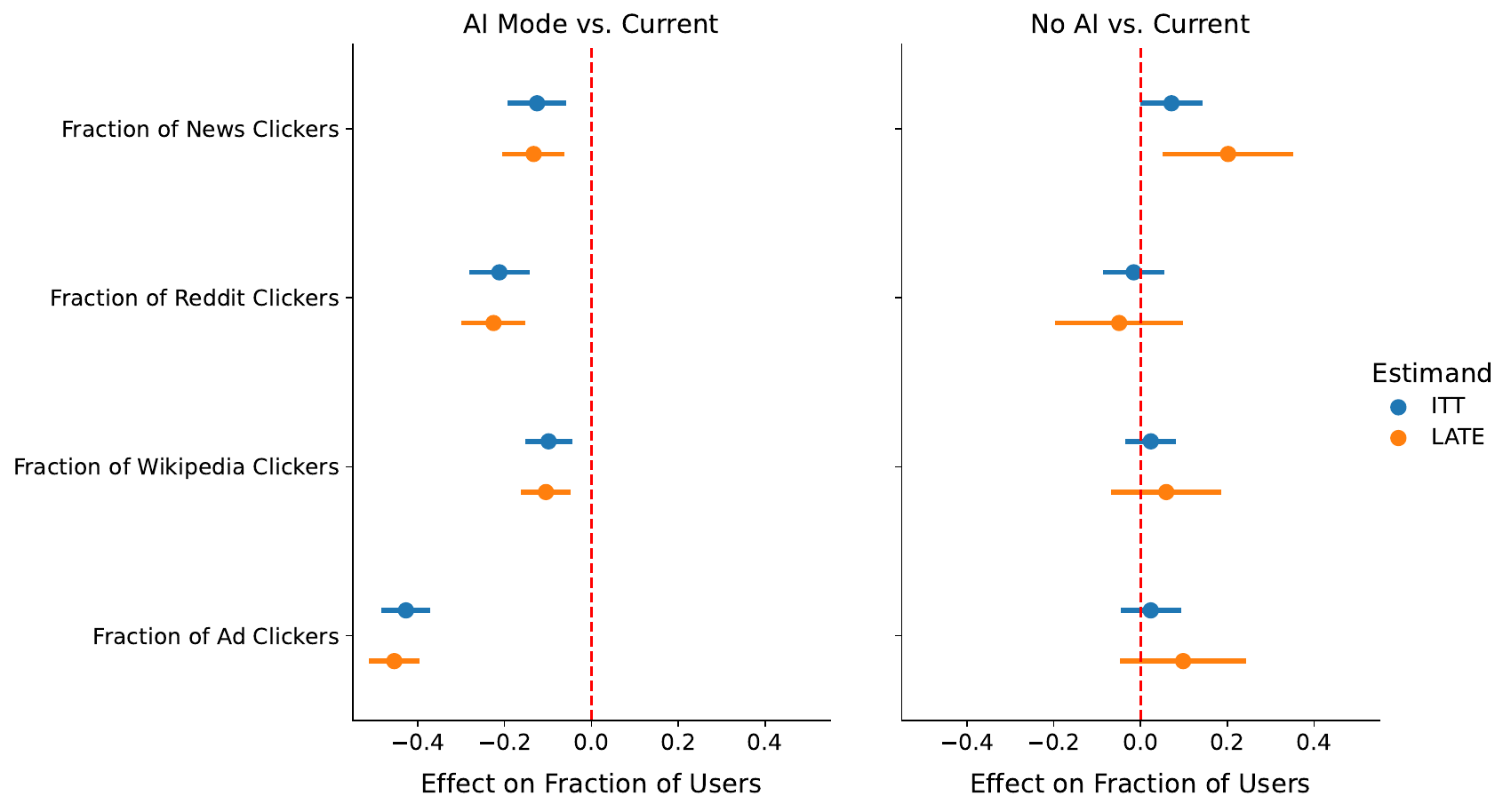} 
    % Captions go below figures
    \caption{\textbf{Effects of AI Mode Search and No AI Search on Secondary Click Outcomes.} Effects on Clicks per Day not shown because other click outcomes are on the same, interpretable scale.}
    \label{fig:secondary_clicks} % give each figure a logical label name
\end{figure}

\begin{figure}[ht]
	\centering
    \includegraphics[width=\textwidth]{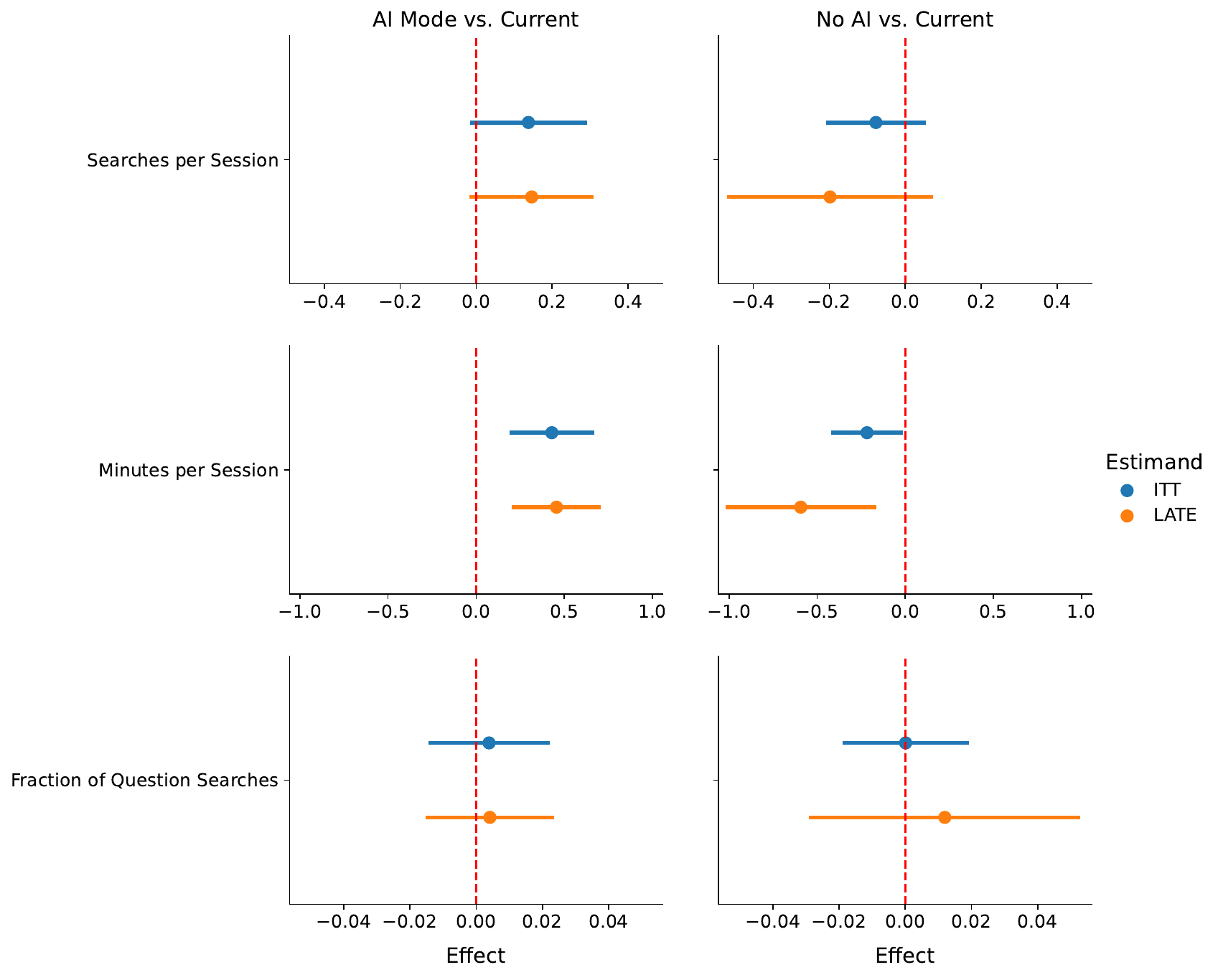}
    \caption{\textbf{Effects of AI Mode Search and No AI Search on Secondary Search Outcomes.}}
    \label{fig:secondary_search}
\end{figure}

\paragraph{Clicks to Specific Domains (H5)} Assignment to AI Mode Search significantly reduced the fraction of users clicking through to news sites (-12.5pp [95\% CI: -18.7, -6.3; $p< 0.001$]), Reddit (-21.2pp [95\% CI: -27.5, -14.8; $p< 0.001$]), Wikipedia (-9.9pp [95\% CI: -14.7, -5.0; $p< 0.001$]). The fraction of users clicking on ads decreased by -42.7pp because AI Mode did not surface ads at the time of our experiment. These results are consistent with H5a (Figure \ref{fig:secondary_clicks} and Table \ref{tab:secondary_clicks}), and are robust to using clicks per day instead of fraction of users clicking (Figure \ref{fig:robustness_clicks}) and to an alternative classification of news domains (Table \ref{tab:robustness_news_def}). In partial support of H5b, exposure to No AI Search increased the fraction of users clicking through to news sites, though we find no evidence of impacts on Reddit, Wikipedia, or ad clicks; the news-site effect is not robust to using clicks per day or an alternative domain classification.
\paragraph{Session Characteristics (H6)} Assignment to AI Mode Search significantly increased minutes per session by 0.43 minutes [95\% CI: 0.20, 0.66; $p = 0.001$], consistent with H6a (Figure \ref{fig:secondary_search} and Table \ref{tab:secondary_search}). This is corroborated by our session-level robustness check (Figure \ref{fig:session_level}, left panel). We do not find evidence of impacts on searches per session (H6c) or the fraction of queries with question words (i.e., ``who'', ``what'', ``where'', ``why'', ``when'', or ``how''; H6a). Exposure to No AI Search decreased minutes per session by -0.59 minutes [95\% CI: -1.01, -0.18; $p = 0.015$], consistent with H6b; the session-level estimate is directionally consistent but not statistically significant. We find no evidence of impacts of No AI Search on searches per session (H6d) or question-form queries (H6b).

\begin{figure}[ht]
	\centering
    \includegraphics[width=\textwidth]{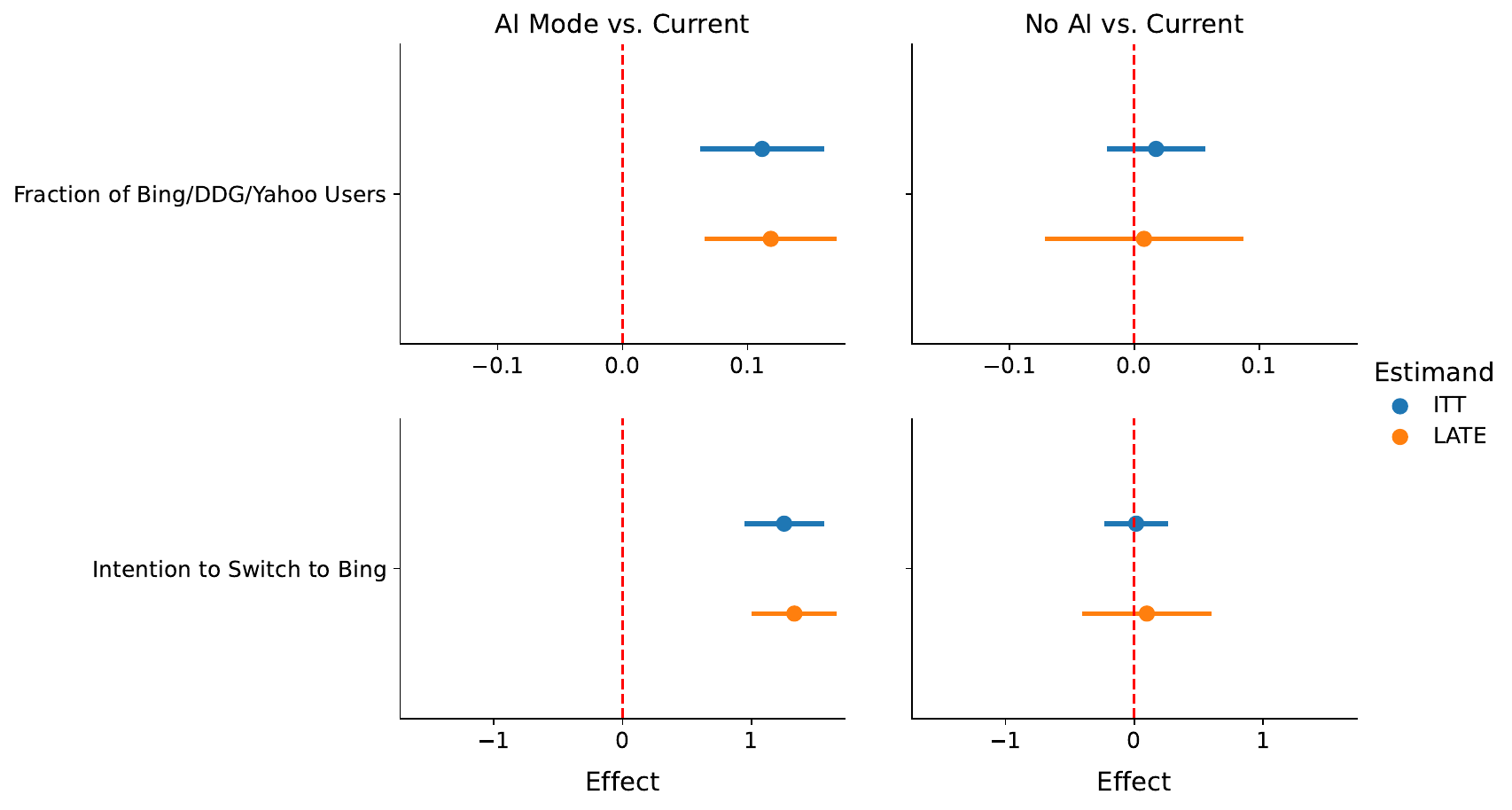}
    \caption{\textbf{Effects of AI Mode Search and No AI Search on Secondary Substitution Outcomes.}}
    \label{fig:secondary_substitution}
\end{figure}
\paragraph{Search Engine Substitution (H7)} Assignment to AI Mode Search significantly increased the fraction of users searching on a competitor engine (Bing, DuckDuckGo, or Yahoo) by 11.2pp [95\% CI: 6.4, 16.0; $p< 0.001$] and intention to switch to Bing by 1.25 points on a 7-point scale [95\% CI: 0.96, 1.55; $p< 0.001$], consistent with H7a (Figure \ref{fig:secondary_substitution} and Table \ref{tab:secondary_substitution}). Consistent with H7b, we do not find evidence that exposure to No AI Search impacted substitution to competitor engines or long-term intention to switch.

\subsection*{Effects on User Perceptions}
\begin{figure}[htp]
	\centering
    \includegraphics[width=\textwidth]{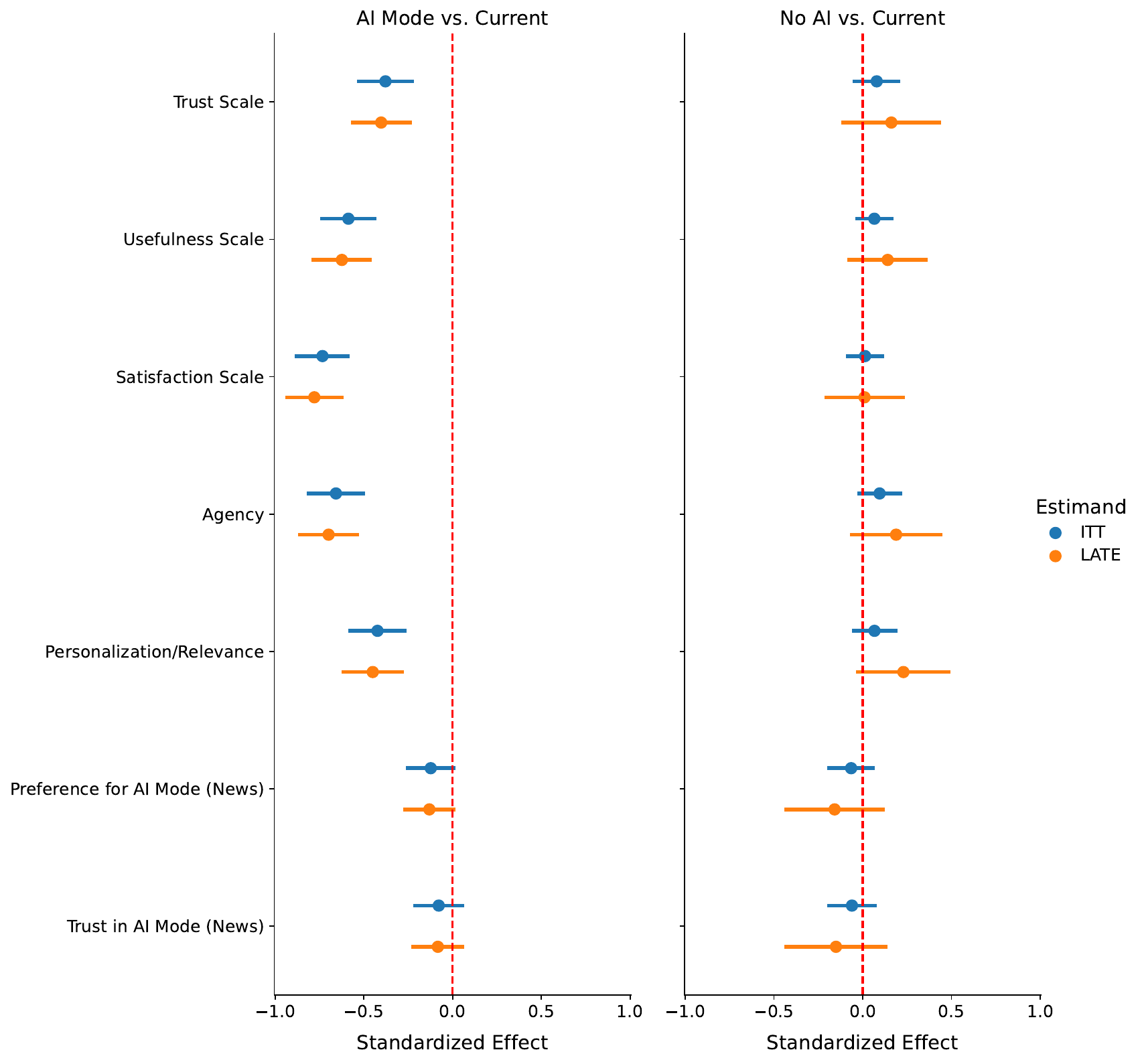} 
    \caption{\textbf{Effects of AI Mode Search and No AI Search on Secondary Survey Outcomes.} Trust Scale shown for reference. Effects are presented in standardized units because some scales have 7 points (trust, agency, and personalization/relevance), some scales have 5 points (usefulness, satisfaction), and some questions are binary (news comparisons).}
    \label{fig:secondary_survey} 
\end{figure}

\paragraph{Trust (H1)} Assignment to AI Mode Search significantly decreased trust in information on Google by -0.34 points on a 7-point scale [95\% CI: -0.47, -0.20; $p< 0.001$], consistent with H1a (Figure \ref{fig:primary} and Table \ref{tab:primary}). We do not find evidence that exposure to No AI Search impacted trust [95\% CI: -0.10, 0.38; $p = 0.107$], failing to support H1b.

\paragraph{Satisfaction, Usefulness, and Agency (H4)} Assignment to AI Mode Search significantly reduced perceived usefulness (-0.59 sd [95\% CI: -0.74, -0.44; $p< 0.001$]), satisfaction (-0.73 sd [95\% CI: -0.88, -0.59; $p< 0.001$]), and agency (-0.66 sd [95\% CI: -0.81, -0.51; $p< 0.001$]), resolving H4a in the direction of negative perceptions of AI Mode Search (Figure \ref{fig:secondary_survey} and Table \ref{tab:secondary_survey}). We do not find evidence that exposure to No AI Search impacted any of these outcomes (H4b).

\paragraph{Personalization and Relevance (H4)} Assignment to AI Mode Search significantly decreased perceived personalization/relevance of responses by -0.42 sd [95\% CI: -0.58, -0.27; $p< 0.001$], contrary to H4c, which predicted an increase. We do not find evidence that exposure to No AI Search impacted this outcome, failing to support H4d.

\paragraph{News Context (H4)} We do not find evidence that assignment to AI Mode Search significantly impacted preference for AI Mode in the head-to-head news comparison [95\% CI: -0.25, 0.01; $p = 0.102$] or trust in AI Mode in the news context [95\% CI: -0.21, 0.05; $p = 0.241$], only partially supporting H4e. Consistent with H4f, exposure to No AI Search had no detectable impact on these outcomes.

\subsection*{Heterogeneous Treatment Effects}

We test for heterogeneous effects on our primary outcomes across two pre-registered moderators: (1) pre-treatment familiarity with LLMs and (2) pre-treatment number of Google Search sessions. After correcting for multiple comparisons, only one moderator-outcome pair is significant: AI Mode reduces the number of sessions per day (H3) significantly more for heavy Google users (i.e., more than 4 daily sessions) (estimate = -2.01 sessions per day, $p = 0.005$).

\subsection*{Qualitative Feedback about AI Mode}

% In the post-experiment survey, participants in the AI Mode condition responded to an open-ended question asking about their overall impression of AI Mode. Expressed sentiment towards AI Mode was heterogeneous, with negative responses predominating (33.6\%), followed by positive (29.3\%). Specific codes point to several drivers of negative experience: a loss of control and agency (17.6\%), difficulty navigating to specific websites (15.3\%), and limited links and/or source diversity (13.4\%). Positive experiences included efficiency and time-saving (14.0\%) and faithful summarization (6.2\%).

At the end of the study, in the post-experiment survey, participants in the AI Mode condition responded to the open-ended question: ``What was your overall impression of Google's AI Mode after one week of using it?''. We qualitatively coded each response (N=309), assigning each response 1--4 codes to describe its theme. We developed codes inductively, continuing until saturation was reached at 17 distinct codes across six dimensions: overall sentiment, search preference, usability, information quality, trust and credibility, and autonomy and control.
% The codes we arrived at were: positive, negative, mixed, prefers traditional search, prefers AI search, limited source diversity, bad at navigational queries, bugs and UI issues, hallucination and accuracy concern, efficient and time saving, too verbose response, faithful AI summarization, information retrieval failure, distrust AI responses, loss of control and user agency, adaptation and learning curve.

Expressed sentiment towards AI Mode was heterogeneous, with negative responses predominating (33.6\%, $n=103$), followed by positive (29.3\%, $n=90$); mixed responses, in which participants expressed both favorable and unfavorable views, comprised 13.7\% of responses ($n=42$).

Specific codes point to several drivers of negative experience. Usability issues were prominent: participants expressed a loss of control and user agency (17.6\%, $n=54$); as one participant noted, ``I like to choose when I use AI''. Difficulty navigating to specific websites was also common (15.3\%, $n=47$) with one participant describing how, ``when I just wanted to land on the webpage I would type in, it would tell me about the website instead''. Limited links and/or source diversity was noted in 13.4\% of responses ($n=41$). Participants also noted that AI outputs were too verbose (6.2\%, $n=19$; e.g., ``it gives way too much info for easy searches'') and raised hallucination and accuracy concerns (4.6\%, $n=14$; e.g., ``the AI mode provided a lot of inaccurate and inconsistent information''). An explicit preference for traditional search, and UI bugs also appeared but less frequently.

Positive experiences included efficiency and time saving (14.0\%, $n=43$; e.g., ``the AI does a strong job of pulling exactly what I was looking for, saving time'') and faithful summarization (6.2\%, $n=19$). Fewer participants expressed trust in AI results (1.6\%, n=5), a preference for AI summarization over traditional search (1.6\%, $n=5$), or adaptation to AI Mode over time (3.6\%, $n=11$; e.g., ``at first the AI Mode threw me off, but I got used to it quickly'').

\section{Discussion}
Our study provides causal evidence about the effects of integrating generative AI into web search. Leveraging an in-situ browser experiment that manipulated users' access to AI Overviews and AI Mode in Google Search over a 10-day study period, we show that the current trajectory of AI-mediated search presents a clear tradeoff: as Google increasingly answers users' questions directly, referral traffic to the broader web declines substantially, yet these changes do not currently yield commensurate improvements in users' trust or overall search experience.

% \paragraph{AI in Search reduces publisher traffic}
At the time of our study, AI Overviews had already become a standard component of Google Search. Removing AI Overviews increased external click-through by 8.8 percentage points, with suggestive evidence of even greater referral traffic to news publishers.

Looking ahead, Google positions AI Mode as the future of search, and its effects on traffic were greater and broader across domains: relative to standard Google Search, enforcing AI Mode reduced external click-through by 18.8 percentage points, reduced daily search sessions, and reduced the fraction of users clicking through to news sites, Reddit, and Wikipedia alike (H5a). Notably, AI Mode increased minutes per session even as it reduced click-through and search sessions per day, showing users spent more time within the AI-mediated search experience while visiting the broader web less often. This reduction in search sessions was significantly larger among heavy Google users. 

Our findings about impacts on click-through rates are consistent with publisher-reported traffic declines since AI features launched~\cite{simonetti_news_2025, chapekis2025google} and observational evidence that AI Overviews reduced Wikipedia traffic by 15\%~\cite{khosravi2026impact}, and our experimental estimates suggest the impact generalizes well beyond Wikipedia and news. Taken together, these results suggest that AI Mode's design, at least as implemented at the time of our study, may compound the traffic pressures of AI Overviews that we document here and that publishers have separately reported, with direct consequences for revenue models, including advertising models~\cite{weide_how_2025}, that are dependent on traffic.

% \paragraph{AI in Search does not improve, and may worsen, the user experience} 
Despite these significant effects on traffic, we find few corresponding gains for users. Removing AI Overviews produced no detectable change in perceived trust in information found on Google, nor on measures of search experience, suggesting that benefits of AI Overviews are not reflected in these outcomes. AI Mode's effects, relative to current Google Search, decreased perceived trust in information on Google, usefulness, satisfaction, agency and relevance, and increased substitution to competing search engines. Qualitative responses help explain these effects: although some participants appreciated faster access to answers, they more frequently described feeling less in control of their search process, encountered difficulty in reaching desired sites, and saw fewer and less diverse sources. Taken together, these results suggest that in its current form, conversational AI search is viewed less favorably by both publishers and searchers. 

% \paragraph{An early snapshot of Google's conversational AI search}
Our estimates reflect an early snapshot of Google's AI Mode: at the time of our study, AI Mode did not yet surface ads or rely on other traditional ad-revenue mechanisms. How Google chooses to monetize AI Mode will be worth monitoring, particularly given the publisher dynamics we document. As they inevitably do so --- through advertising, sponsored results, or other mechanisms --- we anticipate user satisfaction and other measures will shift away from this pre-monetization baseline.

Our results also point to a tension for Google. AI Mode is designed to help Google compete with other AI chatbot products like ChatGPT for the growing share of queries handled through conversational interfaces~\cite{gottfried_americans_2026}, which also serves Google's interest in keeping users on its site. Yet many users remain unconvinced. Our findings showed that AI Mode drove significant substitution toward competing search engines (an 11.2\% increase in competitor search use, H7a), along with substantial intent among these users to switch to Bing if AI Mode were required over the longer term. If users remain firm in their mistrust and dislike of AI Mode, Google will face a tension between pushing the product on users (to acclimate them as much as for its other benefits), and risking alienation that could cost it ground to competitors in a search market that today is nearly monopolistic. 
% usability and trust issues persist, it raises the question of how Google will consolidate its search product in the face of competition from both traditional search engines and AI chatbots.

% \paragraph{Implications for publishers, the information ecosystem, and the future of search}
Importantly, the introduction of AI features in search shift Google away from serving as an intermediary that connects users to external sources (the canonical definition of a search engine) and into a destination for answers unto itself. This shift already carries regulatory consequences: a recent German court ruling held that Google's AI Overviews cause it to function as a content \textit{provider} rather than a neutral intermediary, on the grounds that ``AI-produced summaries amounted to the company's own content rather than a mere display of third-party information''~\cite{reuters_german_2026}. This determination turns on the same distinction to which our results speak: whether AI in search primarily facilitates users' discovery of third-party publishers, or increasingly substitutes for it.

Given these implications, high stakes for Google, publishers, and users alike, we argue that AI-mediated search should be evaluated not only in terms of the quality of answers generated, but also in terms of how these systems reshape the relationship between users, publishers, and the broader web---from the sustainability of publishers, to Google's status as an intermediary, to users' own right to control their discovery, evaluation, and engagement with with information.
\paragraph{Limitations}
Our study has several limitations, which also point to areas for future work.  

Compliance in the No AI Search condition degraded over the study period due to a Google interface change, falling to roughly 50\% overall. Therefore, we report LATE estimates (Section~\ref{results:compliance-attrition}), which rely on the assumption that assignment only affected outcomes through AI Overview exposure (i.e., the exclusion restriction). If the interface change altered participants' experience more broadly, that assumption may not hold.

We do not find evidence of differential attrition by condition, which supports treating our AI Mode estimates as causal rather than descriptive (Section~\ref{results:compliance-attrition}); we note, however, that this does not rule out the possibility that unfamiliarity with the interface affected participant behavior in other ways.

Our sample was recruited from Prolific and a university work-study program, and skewed younger, more highly educated, and more left-leaning than the general US population (Figure~\ref{fig:participant_demographics}); effects may differ in magnitude or direction in a more representative sample.   

Our experiment spanned seven days of treatment exposure, which allows us to detect short-term behavioral and attitudinal shifts but not longer-run adaptation---it is possible that some effects would attenuate, or that new ones would emerge, over a longer exposure period; future work should examine the impacts of longer-term adjustment on behavior and beliefs. Baseline AI Mode usage was minimal (0.6\% of searches) prior to our intervention, meaning our AI Mode condition reflects a forced, full-adoption scenario rather than the gradual, opt-in uptake most users would experience in practice; effects of organic AI Mode adoption may differ from the effects of the enforced assignment we study here.   

Our design intentionally measured effects across the distribution of real-world search tasks; future work should measure effects for specific high-stakes contexts and search tasks (e.g., health, legal, or financial information-seeking), where the consequences of reduced trust or click-through may differ. Finally, we were not powered to detect small effects on secondary click and search outcomes, and null effects on these outcomes should not be interpreted as evidence of no effect. The same caution applies to secondary survey outcomes, which have fairly wide confidence intervals.
\section{Acknowledgments}
We thank David Lazer, Emma Lurie, and Ro Encarnación for helpful feedback and discussions. We thank all of our pilot and full study participants.

Generative AI (GPT-5.5-mini) was used to lightly edit and rephrase author-written text for clarity and fluency. All analyses, interpretations, and conclusions were developed and verified by the authors.

\section{Competing interest}
The authors declare that they have no competing interests.

\section{Funding}
This work was supported in part by the National Science Foundation under award IIS-2442711. J.G. was supported by the Platforms for Exchange and Allocation of Resources (PEAR) program, a National Science Foundation Research Traineeship.

\section{Author contributions}

Conceptualization: S.W., J.G., Y.B., C.W., D.M., Methodology: S.W., J.G., Y.B., C.W., D.M., Software: S.W., J.G., Validation: S.W., J.G., Formal analysis: S.W., J.G., Investigation: S.W., J.G., Resources: S.W., J.G., Data curation: S.W., J.G., Writing - original draft: S.W., J.G.; Writing - review \& editing: S.W., J.G., Y.B., C.W., D.M.; Supervision: Y.B., C.W., D.M.; Project administration: S.W., J.G.; Funding acquisition: S.W., J.G., Y.B., C.W., D.M.

\section{Data availability}
The repository with aggregated data and code necessary to reproduce the results in this paper are available on Harvard Dataverse~\cite{dataverse_gleason}.
The preregistration of the study is available on OSF~\cite{osf_anonymous}.

\section{Ethics statement}
The study was approved by the Institutional Review Boards of both participating universities. All participants provided informed consent prior to installing the browser extension and enrolling in the study. Consent materials outlined the types of data to be collected and stated that participants might experience modifications to their Google Search interface during the study period. Participants were informed that browsing history and HTML snapshots of Google Search result pages would be collected solely for research purposes, stored securely, accessible only to the research team, and anonymized prior to analysis. Participation was entirely voluntary. Participants could withdraw consent, discontinue participation, or request deletion of their data at any time. 

To protect participant privacy, we stripped all collected data of identifiers where possible. However, because browsing data lacks a fixed structure, it is not possible to systematically remove all identifying information; collected URLs may contain embedded identifiers, and HTML snapshots may include personal details such as names or email addresses appearing in dynamic locations. Given these limitations, raw browsing data and HTML snapshots will not be publicly released.  

At the conclusion of the study, participants were financially compensated through the recruitment platform and debriefed via a statement shown in the web app after submitting the post-survey. The debriefing explained the study's purpose, details about the three conditions, and the rationale for withholding this information from participants in order to minimize demand effects.

\bibliographystyle{unsrt}
\bibliography{reference}

@article{huang2026answer,
  title={Answer Bubbles: Information Exposure in AI-Mediated Search},
  author={Huang, Michelle and Goyal, Agam and Saha, Koustuv and Chandrasekharan, Eshwar},
  journal={arXiv preprint arXiv:2603.16138},
  year={2026}
}

@techreport{lin2016green,
  title     = {Standard Operating Procedures for {Don Green's} Lab at {Columbia}},
  author    = {Lin, Winston and Green, Donald P. and Coppock, Alexander},
  year      = {2016},
  month     = {June},
  note      = {Version 1.05},
  url       = {https://alexandercoppock.com/Green-Lab-SOP/Green_Lab_SOP.pdf}
}

@article{anderson2008multiple,
  title={Multiple inference and gender differences in the effects of early intervention: A reevaluation of the Abecedarian, Perry Preschool, and Early Training Projects},
  author={Anderson, Michael L},
  journal={Journal of the American statistical Association},
  volume={103},
  number={484},
  pages={1481--1495},
  year={2008},
  publisher={Taylor \& Francis}
}

@dataset{yang2025newsdomains,
  author       = {Yang, Kai-Cheng},
  title        = {A list of news domains},
  month        = mar,
  year         = {2025},
  publisher    = {Zenodo},
  doi          = {10.5281/zenodo.15073918},
  url          = {https://doi.org/10.5281/zenodo.15073918},
}

@misc{reid_generative_2024,
	title = {Generative {AI} in {Search}: {Let} {Google} do the searching for you},
	shorttitle = {Generative {AI} in {Search}},
	url = {https://blog.google/products-and-platforms/products/search/generative-ai-google-search-may-2024/},
	language = {en-us},
	urldate = {2026-02-27},
	journal = {Google},
	author = {Reid, Elizabeth},
	month = may,
	year = {2024},
}

@inproceedings{kim_im_2024,
	address = {New York, NY, USA},
	series = {{FAccT} '24},
	title = {"{I}'m {Not} {Sure}, {But}...": {Examining} the {Impact} of {Large} {Language} {Models}' {Uncertainty} {Expression} on {User} {Reliance} and {Trust}},
	isbn = {979-8-4007-0450-5},
	shorttitle = {"{I}'m {Not} {Sure}, {But}..."},
	url = {https://dl.acm.org/doi/10.1145/3630106.3658941},
	doi = {10.1145/3630106.3658941},
	urldate = {2026-02-09},
	booktitle = {Proceedings of the 2024 {ACM} {Conference} on {Fairness}, {Accountability}, and {Transparency}},
	publisher = {Association for Computing Machinery},
	author = {Kim, Sunnie S. Y. and Liao, Q. Vera and Vorvoreanu, Mihaela and Ballard, Stephanie and Vaughan, Jennifer Wortman},
	month = jun,
	year = {2024},
	pages = {822--835},
}

@inproceedings{kim_fostering_2025,
	address = {New York, NY, USA},
	series = {{CHI} '25},
	title = {Fostering {Appropriate} {Reliance} on {Large} {Language} {Models}: {The} {Role} of {Explanations}, {Sources}, and {Inconsistencies}},
	isbn = {979-8-4007-1394-1},
	shorttitle = {Fostering {Appropriate} {Reliance} on {Large} {Language} {Models}},
	url = {https://dl.acm.org/doi/10.1145/3706598.3714020},
	doi = {10.1145/3706598.3714020},
	urldate = {2025-10-29},
	booktitle = {Proceedings of the 2025 {CHI} {Conference} on {Human} {Factors} in {Computing} {Systems}},
	publisher = {Association for Computing Machinery},
	author = {Kim, Sunnie S. Y. and Vaughan, Jennifer Wortman and Liao, Q. Vera and Lombrozo, Tania and Russakovsky, Olga},
	month = apr,
	year = {2025},
	pages = {1--19},
}

@article{simonetti_news_2025,
	chapter = {Business},
	title = {News {Sites} {Are} {Getting} {Crushed} by {Google}’s {New} {AI} {Tools}},
	issn = {0099-9660},
	url = {https://www.wsj.com/tech/ai/google-ai-news-publishers-7e687141},
	language = {en-US},
	urldate = {2025-10-20},
	journal = {Wall Street Journal},
	author = {Simonetti, Isabella and Blunt, Katherine},
	month = jun,
	year = {2025},
}

@misc{williams_why_2024,
	title = {Why {Google}’s {AI} {Overviews} gets things wrong},
	url = {https://www.technologyreview.com/2024/05/31/1093019/why-are-googles-ai-overviews-results-so-bad/},
	month = may,
	year = {2024},
	language = {en},
	journal = {MIT Technology Review},
	author = {Williams, Rhiannon},
}

@misc{bearne_publishers_2025,
	title = {Publishers fear {AI} summaries are hitting online traffic},
    author = {Bearne, Suzanne},
	url = {https://www.bbc.com/news/articles/c0mlvryx0exo},
	urldate = {2025-09-26},
    year = {2025},
}

@misc{li_human_2025,
	title = {Human {Trust} in {AI} {Search}: {A} {Large}-{Scale} {Experiment}},
	shorttitle = {Human {Trust} in {AI} {Search}},
	url = {http://arxiv.org/abs/2504.06435},
	doi = {10.48550/arXiv.2504.06435},
	urldate = {2025-10-15},
	publisher = {arXiv},
	author = {Li, Haiwen and Aral, Sinan},
	month = apr,
	year = {2025},
	note = {arXiv:2504.06435 [cs]},
}

@misc{chen_generative_2025,
	title = {Generative {Engine} {Optimization}: {How} to {Dominate} {AI} {Search}},
	shorttitle = {Generative {Engine} {Optimization}},
	url = {http://arxiv.org/abs/2509.08919},
	doi = {10.48550/arXiv.2509.08919},
	urldate = {2025-10-13},
	publisher = {arXiv},
	author = {Chen, Mahe and Wang, Xiaoxuan and Chen, Kaiwen and Koudas, Nick},
	month = sep,
	year = {2025},
	note = {arXiv:2509.08919 [cs]},
}

@article{shah_envisioning_2024,
	title = {Envisioning {Information} {Access} {Systems}: {What} {Makes} for {Good} {Tools} and a {Healthy} {Web}?},
	volume = {18},
	issn = {1559-1131},
	shorttitle = {Envisioning {Information} {Access} {Systems}},
	url = {https://dl.acm.org/doi/10.1145/3649468},
	doi = {10.1145/3649468},
	number = {3},
	urldate = {2025-10-09},
	journal = {ACM Trans. Web},
	author = {Shah, Chirag and Bender, Emily M.},
	month = apr,
	year = {2024},
	pages = {33:1--33:24},
}

@inproceedings{liu_evaluating_2023,
  title={Evaluating verifiability in generative search engines},
  author={Liu, Nelson F and Zhang, Tianyi and Liang, Percy},
  booktitle={Findings of the Association for Computational Linguistics: EMNLP 2023},
  pages={7001--7025},
  year={2023}
}

@inproceedings{narayanan_venkit_search_2025,
	address = {Athens Greece},
	title = {Search {Engines} in the {AI} {Era}: {A} {Qualitative} {Understanding} to the {False} {Promise} of {Factual} and {Verifiable} {Source}-{Cited} {Responses} in {LLM}-based {Search}},
	shorttitle = {Search {Engines} in the {AI} {Era}},
	url = {https://dl.acm.org/doi/10.1145/3715275.3732089},
	doi = {10.1145/3715275.3732089},
	urldate = {2025-09-30},
	booktitle = {Proceedings of the 2025 {ACM} {Conference} on {Fairness}, {Accountability}, and {Transparency}},
	publisher = {ACM},
	author = {Narayanan Venkit, Pranav and Laban, Philippe and Zhou, Yilun and Mao, Yixin and Wu, Chien-Sheng},
	month = jun,
	year = {2025},
	pages = {1325--1340},
}

@inproceedings{spatharioti_effects_2025,
	address = {Yokohama Japan},
	title = {Effects of {LLM}-based {Search} on {Decision} {Making}: {Speed}, {Accuracy}, and {Overreliance}},
	copyright = {https://creativecommons.org/licenses/by-nc-sa/4.0/},
	shorttitle = {Effects of {LLM}-based {Search} on {Decision} {Making}},
	url = {https://dl.acm.org/doi/10.1145/3706598.3714082},
	doi = {10.1145/3706598.3714082},
	urldate = {2025-09-30},
	booktitle = {Proceedings of the 2025 {CHI} {Conference} on {Human} {Factors} in {Computing} {Systems}},
	publisher = {ACM},
	author = {Spatharioti, Sofia Eleni and Rothschild, David and Goldstein, Daniel G and Hofman, Jake M},
	month = apr,
	year = {2025},
	pages = {1--15},
}

@inproceedings{sharma_generative_2024,
	address = {Honolulu HI USA},
	title = {Generative {Echo} {Chamber}? {Effect} of {LLM}-{Powered} {Search} {Systems} on {Diverse} {Information} {Seeking}},
	copyright = {https://creativecommons.org/licenses/by/4.0/},
	shorttitle = {Generative {Echo} {Chamber}?},
	url = {https://dl.acm.org/doi/10.1145/3613904.3642459},
	doi = {10.1145/3613904.3642459},
	language = {en},
	urldate = {2025-09-30},
	booktitle = {Proceedings of the {CHI} {Conference} on {Human} {Factors} in {Computing} {Systems}},
	publisher = {ACM},
	author = {Sharma, Nikhil and Liao, Q. Vera and Xiao, Ziang},
	month = may,
	year = {2024},
	pages = {1--17},
}

@inproceedings{lurie_investigating_2018,
	address = {Amsterdam Netherlands},
	title = {Investigating the {Effects} of {Google}'s {Search} {Engine} {Result} {Page} in {Evaluating} the {Credibility} of {Online} {News} {Sources}},
	copyright = {https://www.acm.org/publications/policies/copyright\_policy\#Background},
	url = {https://dl.acm.org/doi/10.1145/3201064.3201095},
	doi = {10.1145/3201064.3201095},
	language = {en},
	urldate = {2025-09-29},
	booktitle = {Proceedings of the 10th {ACM} {Conference} on {Web} {Science}},
	publisher = {ACM},
	author = {Lurie, Emma and Mustafaraj, Eni},
	month = may,
	year = {2018},
	pages = {107--116},
}

@article{gleason_google_2023,
	title = {Google the {Gatekeeper}: {How} {Search} {Components} {Affect} {Clicks} and {Attention}},
	volume = {17},
	copyright = {Copyright (c) 2023 Association for the Advancement of Artificial Intelligence},
	issn = {2334-0770},
	shorttitle = {Google the {Gatekeeper}},
	url = {https://ojs.aaai.org/index.php/ICWSM/article/view/22142},
	doi = {10.1609/icwsm.v17i1.22142},
	language = {en},
	urldate = {2025-09-21},
	journal = {Proceedings of the International AAAI Conference on Web and Social Media},
	author = {Gleason, Jeffrey and Hu, Desheng and Robertson, Ronald E. and Wilson, Christo},
	month = jun,
	year = {2023},
	pages = {245--256},
}

@inproceedings{li_good_2009,
	address = {Boston MA USA},
	title = {Good abandonment in mobile and {PC} internet search},
	copyright = {https://www.acm.org/publications/policies/copyright\_policy\#Background},
	url = {https://dl.acm.org/doi/10.1145/1571941.1571951},
	doi = {10.1145/1571941.1571951},
	urldate = {2025-09-22},
	booktitle = {Proceedings of the 32nd international {ACM} {SIGIR} conference on {Research} and development in information retrieval},
	publisher = {ACM},
	author = {Li, Jane and Huffman, Scott and Tokuda, Akihito},
	month = jul,
	year = {2009},
	pages = {43--50},
}

@article{lam_sociotechnical_2023,
	title = {Sociotechnical {Audits}: {Broadening} the {Algorithm} {Auditing} {Lens} to {Investigate} {Targeted} {Advertising}},
	volume = {7},
	shorttitle = {Sociotechnical {Audits}},
	url = {https://dl.acm.org/doi/10.1145/3610209},
	doi = {10.1145/3610209},
	number = {CSCW2},
	urldate = {2023-12-22},
	journal = {Proceedings of the ACM on Human-Computer Interaction},
	author = {Lam, Michelle S. and Pandit, Ayush and Kalicki, Colin H. and Gupta, Rachit and Sahoo, Poonam and Metaxa, Danaë},
	month = oct,
	year = {2023},
	pages = {360:1--360:37},
}

@misc{yang_news_2025,
	title = {News {Source} {Citing} {Patterns} in {AI} {Search} {Systems}},
	url = {http://arxiv.org/abs/2507.05301},
	doi = {10.48550/arXiv.2507.05301},
	urldate = {2026-03-04},
	publisher = {arXiv},
	author = {Yang, Kai-Cheng},
	month = jul,
	year = {2025},
	note = {arXiv:2507.05301 [cs]},
}

@inproceedings{miroyan_search_2026,
  title={Search arena: Analyzing search-augmented llms},
  author={Miroyan, Mihran and Wu, Tsung-Han and King, Logan and Li, Tianle and Pan, Jiayi and Hu, Xinyan and Chiang, Wei-Lin and Angelopoulos, Anastasios and Norouzi, Narges and Gonzalez, Joseph E and others},
  booktitle={International Conference on Learning Representations},
  volume={2026},
  pages={41109--41146},
  year={2026}
}

@misc{padilla_impact_2025,
	address = {Rochester, NY},
	type = {{SSRN} {Scholarly} {Paper}},
	title = {The {Impact} of {LLM} {Adoption} on {Online} {User} {Behavior}},
	url = {https://papers.ssrn.com/abstract=5393256},
	doi = {10.2139/ssrn.5393256},
	language = {en},
	urldate = {2026-03-06},
	publisher = {Social Science Research Network},
	author = {Padilla, Nicolas and Lam, H. Tai and Lambrecht, Anja and Hollenbeck, Brett},
	month = dec,
	year = {2025},
}

@article{wang_lower_2024,
	title = {Lower {Quantity}, {Higher} {Quality}: {Auditing} {News} {Content} and {User} {Perceptions} on {Twitter}/{X} {Algorithmic} versus {Chronological} {Timelines}},
	volume = {8},
	copyright = {https://creativecommons.org/licenses/by-nc-sa/4.0/},
	issn = {2573-0142},
	shorttitle = {Lower {Quantity}, {Higher} {Quality}},
	url = {https://dl.acm.org/doi/10.1145/3687046},
	doi = {10.1145/3687046},
	language = {en},
	number = {CSCW2},
	urldate = {2025-08-31},
	journal = {Proceedings of the ACM on Human-Computer Interaction},
	publisher = {Association for Computing Machinery (ACM)},
	author = {Wang, Stephanie and Huang, Shengchun and Zhou, Alvin and Metaxa, Danaë},
	month = nov,
	year = {2024},
	pages = {1--25},
}

@misc{xu_ai_2025,
	title = {{AI} summaries in online search influence users' attitudes},
	copyright = {Creative Commons Attribution 4.0 International},
	url = {https://arxiv.org/abs/2511.22809},
	doi = {10.48550/ARXIV.2511.22809},
	language = {en},
	urldate = {2026-05-06},
	publisher = {arXiv},
	author = {Xu, Yiwei and Dash, Saloni and Kang, Sungha and Liao, Wang and Spiro, Emma S.},
	year = {2025},
	note = {Version Number: 2},
}

@misc{memon_search_2024,
	title = {Search {Engines} {Post}-{ChatGPT}: {How} {Generative} {Artificial} {Intelligence} {Could} {Make} {Search} {Less} {Reliable}},
	shorttitle = {Search {Engines} {Post}-{ChatGPT}},
	url = {http://arxiv.org/abs/2402.11707},
	doi = {10.48550/arXiv.2402.11707},
	urldate = {2026-05-06},
	publisher = {arXiv},
	author = {Memon, Shahan Ali and West, Jevin D.},
	month = feb,
	year = {2024},
	note = {arXiv:2402.11707 [cs]},
}

@misc{xu_chatgpt_2023,
	title = {{ChatGPT} vs. {Google}: {A} {Comparative} {Study} of {Search} {Performance} and {User} {Experience}},
	shorttitle = {{ChatGPT} vs. {Google}},
	url = {http://arxiv.org/abs/2307.01135},
	doi = {10.48550/arXiv.2307.01135},
	urldate = {2026-05-06},
	publisher = {arXiv},
	author = {Xu, Ruiyun and Feng, Yue and Chen, Hailiang},
	month = jul,
	year = {2023},
	note = {arXiv:2307.01135 [cs]},
}

@misc{osf_anonymous,
  author = {Anonymous},
  title = {Impact of AI-Driven Search on User Behaviors and Perceptions},
  year = {2026},
  url = {https://osf.io/s8xnc/overview?view_only=79295266bf274da3b5f5fed6532796c8},
  organization = {OSF},
}

@article{flanagin2000perceptions,
  title={Perceptions of Internet information credibility},
  author={Flanagin, Andrew J and Metzger, Miriam J},
  journal={Journalism \& mass communication quarterly},
  volume={77},
  number={3},
  pages={515--540},
  year={2000},
  publisher={SAGE Publications Sage CA: Los Angeles, CA}
}

@article{van1997simple,
  title={A simple procedure for the assessment of acceptance of advanced transport telematics},
  author={Van Der Laan, Jinke D and Heino, Adriaan and De Waard, Dick},
  journal={Transportation Research Part C: Emerging Technologies},
  volume={5},
  number={1},
  pages={1--10},
  year={1997},
  publisher={Elsevier}
}

@misc{google_gclid,
  author       = {Google},
  title        = {Google Click Identifier ({GCLID}): Definition},
  howpublished = {Google Ads Help},
  url          = {https://support.google.com/google-ads/answer/9744275?hl=en},
  note         = {Accessed: 2026-05-11}
}

@misc{chapekis2025google,
  author       = {Chapekis, Athena and Lieb, Anna},
  title        = {Google users are less likely to click on links when an AI summary appears in the results},
  howpublished = {Pew Research Center},
  year         = {2025},
  month        = jul,
  day          = {22},
  url          = {https://www.pewresearch.org/short-reads/2025/07/22/google-users-are-less-likely-to-click-on-links-when-an-ai-summary-appears-in-the-results/},
  note         = {Accessed: 2026-05-11}
}

@misc{reid_ai_2025,
	title = {{AI} in {Search} is driving more queries and higher quality clicks},
	url = {https://blog.google/products/search/ai-search-driving-more-queries-higher-quality-clicks/},
	language = {en-us},
	urldate = {2025-10-20},
	journal = {Google},
	month = aug,
	year = {2025},
    author = {Reid, Liz},
}

@misc{google_personal_intelligence_search,
  author       = {{Google}},
  title        = {Connect your {Google} content apps to get a {Search} experience that's tailored to you},
  howpublished = {Google Search Help Center},
  url          = {https://support.google.com/websearch/?p=PersonalIntelligenceSearch},
  note         = {Accessed: 2026-05-11}
}

@article{khosravi2026impact,
  title={Impact of AI Search Summaries on Website Traffic: Evidence from Google AI Overviews and Wikipedia},
  author={Khosravi, Mehrzad and Yoganarasimhan, Hema},
  journal={Available at SSRN 6164926},
  year={2026}
}

@article{gholami2026beyond,
  title={Beyond search: LLM adoption and web traffic concentration},
  author={Gholami, Samira and Firullo, Cristiana and Cheyre, Cristobal and Acquisti, Alessandro},
  journal={Available at SSRN 6238578},
  year={2026}
}

@misc{stein_expanding_2025,
	title = {Expanding {AI} {Overviews} and introducing {AI} {Mode}},
    author = {Stein, Robby},
	url = {https://blog.google/products-and-platforms/products/search/ai-mode-search/},
	language = {en-us},
	urldate = {2026-05-12},
	journal = {Google},
	month = mar,
	year = {2025},
}

@article{piccardi2025reranking,
  title={Reranking partisan animosity in algorithmic social media feeds alters affective polarization},
  author={Piccardi, Tiziano and Saveski, Martin and Jia, Chenyan and Hancock, Jeffrey and Tsai, Jeanne L and Bernstein, Michael S},
  journal={Science},
  volume={390},
  number={6776},
  pages={eadu5584},
  year={2025},
  publisher={American Association for the Advancement of Science}
}

@inproceedings{diriye_leaving_2012,
	address = {Maui Hawaii USA},
	title = {Leaving so soon?: understanding and predicting web search abandonment rationales},
	isbn = {978-1-4503-1156-4},
	shorttitle = {Leaving so soon?},
	url = {https://dl.acm.org/doi/10.1145/2396761.2398399},
	doi = {10.1145/2396761.2398399},
	language = {en},
	urldate = {2026-06-05},
	booktitle = {Proceedings of the 21st {ACM} international conference on {Information} and knowledge management},
	publisher = {ACM},
	author = {Diriye, Abdigani and White, Ryen and Buscher, Georg and Dumais, Susan},
	month = oct,
	year = {2012},
	pages = {1025--1034},
}

@article{pape_is_2026,
	title = {Is {Competition} {Only} {One} {Click} {Away}? {The} {Digital} {Markets} {Act}’s {Impact} on {Google} {Maps}},
	volume = {45},
	issn = {0732-2399},
	shorttitle = {Is {Competition} {Only} {One} {Click} {Away}?},
	url = {https://pubsonline.informs.org/doi/10.1287/mksc.2025.0159},
	doi = {10.1287/mksc.2025.0159},
	number = {3},
	urldate = {2026-06-17},
	journal = {Marketing Science},
	publisher = {INFORMS},
	author = {Pape, Louis-Daniel and Rossi, Michelangelo},
	month = may,
	year = {2026},
	pages = {596--613},
}

@article{burtch_consequences_2024,
	title = {The consequences of generative {AI} for online knowledge communities},
	volume = {14},
	copyright = {2024 The Author(s)},
	issn = {2045-2322},
	url = {https://www.nature.com/articles/s41598-024-61221-0},
	doi = {10.1038/s41598-024-61221-0},
	language = {en},
	number = {1},
	urldate = {2026-06-17},
	journal = {Scientific Reports},
	publisher = {Nature Publishing Group},
	author = {Burtch, Gordon and Lee, Dokyun and Chen, Zhichen},
	month = may,
	year = {2024},
	pages = {10413},
}

@inproceedings{lyu_wikipedia_2025,
  title={Wikipedia contributions in the wake of chatgpt},
  author={Lyu, Liang and Siderius, James and Li, Hannah and Acemoglu, Daron and Huttenlocher, Daniel and Ozdaglar, Asuman},
  booktitle={Companion Proceedings of the ACM on Web Conference 2025},
  pages={1176--1179},
  year={2025}
}

@article{del_rio-chanona_large_2024,
	title = {Large language models reduce public knowledge sharing on online {Q}\&amp;{A} platforms},
	volume = {3},
	issn = {2752-6542},
	url = {https://doi.org/10.1093/pnasnexus/pgae400},
	doi = {10.1093/pnasnexus/pgae400},
	number = {9},
	journal = {PNAS Nexus},
	author = {del Rio-Chanona, R Maria and Laurentsyeva, Nadzeya and Wachs, Johannes},
	month = sep,
	year = {2024},
	note = {\_eprint: https://academic.oup.com/pnasnexus/article-pdf/3/9/pgae400/59316621/pgae400.pdf},
	pages = {pgae400},
}

@article{hu_auditing_2026,
	title = {Auditing {Google}’s {AI} {Overviews} and {Featured} {Snippets}: {A} {Case} {Study} on {Baby} {Care} and {Pregnancy}},
	volume = {20},
	copyright = {Copyright (c) 2026 Association for the Advancement of Artificial Intelligence},
	issn = {2334-0770},
	shorttitle = {Auditing {Google}’s {AI} {Overviews} and {Featured} {Snippets}},
	url = {https://ojs.aaai.org/index.php/ICWSM/article/view/42681},
	doi = {10.1609/icwsm.v20i1.42681},
	language = {en},
	number = {1},
	urldate = {2026-06-19},
	journal = {Proceedings of the International AAAI Conference on Web and Social Media},
	author = {Hu, Desheng and Baumann, Joachim and Urman, Aleksandra and Lichtenegger, Elsa and Forsberg, Robin and Hannák, Anikó and Wilson, Christo},
	month = may,
	year = {2026},
	pages = {1044--1062},
}

@misc{census_us_2025,
	title = {U.{S}. {Census} {Bureau} {QuickFacts}: {United} {States}},
	shorttitle = {U.{S}. {Census} {Bureau} {QuickFacts}},
	url = {https://www.census.gov/quickfacts/fact/table/US/PST045225},
	language = {en},
	urldate = {2026-06-19},
}

@data{dataverse_gleason,
author = {Gleason, Jeffrey and Wang, Stephanie and Bart, Yakov and Wilson, Christo and Metaxa, Danaé},
publisher = {Harvard Dataverse},
title = {{Replication Data for: ``AI in Search Reduces Traffic to Publishers Without Improving Searcher Experience: A Field Experiment''}},
year = {2026},
version = {V1},
doi = {10.7910/DVN/2O0UCR},
url = {https://doi.org/10.7910/DVN/2O0UCR}
}

@article{conger_google_2026,
	chapter = {Technology},
	title = {Google {Is} {Building} an {A}.{I}. {Fence} {Around} the {Internet} {It} {Once} {Championed}},
	issn = {0362-4331},
	url = {https://www.nytimes.com/2026/07/20/technology/google-ai-open-web.html},
	language = {en-US},
	urldate = {2026-07-27},
	journal = {The New York Times},
	author = {Conger, Kate},
	month = jul,
	year = {2026},
}

@misc{google_ai_2026,
  author       = {{Google}},
  title        = {{AI in Search}},
  year         = {2026},
  howpublished = {\url{https://search.google/ai-in-search/}},
  language = {en-US},
	urldate = {2026-07-27},
}

@article{reuters_german_2026,
	chapter = {Government},
	title = {German media regulator says {Google}'s {AI} {Overviews} subject to {German} media law},
	url = {https://www.reuters.com/legal/government/german-media-regulator-says-googles-ai-overviews-subject-german-media-law-2026-07-14/},
	language = {en},
	urldate = {2026-07-29},
	journal = {Reuters},
	month = jul,
	year = {2026},
}

@misc{weide_how_2025,
	title = {How {AI} {Search} {Threatens} {Publishers} and {AdTech} {Vendors}},
    author = {Weide, Karsten},
    month = {Sep},
    year = {2025},
	url = {https://news.marketecture.tv/p/how-ai-search-threatens-publishers-and-adtech-vendors},
	urldate = {2025-09-29},
}

@misc{gottfried_americans_2026,
	title = {Americans and {AI} 2026: {Chatbots}, {Smart} {Devices} and {Views} on {Impact}},
	shorttitle = {Americans and {AI} 2026},
	url = {https://www.pewresearch.org/internet/2026/06/17/americans-and-ai-2026-chatbots-smart-devices-and-views-on-impact/},
	language = {en-US},
	urldate = {2026-08-02},
	journal = {Pew Research Center},
	author = {Gottfried, Jeffrey and Bishop, William and Anderson, Monica and Faverio, Michelle and Park, Eugenie and McClain, Colleen},
	month = jun,
	year = {2026},
	note = {Section: Artificial Intelligence},
}

% \include{sections/appendix}

%% ---------------------------------------------------------------
%% Supplementary Materials (merged from supplementary.tex)
%% Starts a new page, restarts numbering with an "S" prefix, and
%% applies the SM's caption style, matching the standalone SM file.
%% ---------------------------------------------------------------
\clearpage

\renewcommand{\thetable}{S\arabic{table}}
\renewcommand{\thefigure}{S\arabic{figure}}
\renewcommand{\theequation}{S\arabic{equation}}
\renewcommand{\thesection}{S\arabic{section}}
\setcounter{table}{0}
\setcounter{figure}{0}
\setcounter{equation}{0}
\setcounter{section}{0}

\captionsetup[table]{labelfont=bf, labelsep=period, justification=raggedright, singlelinecheck=false}
\captionsetup[figure]{labelfont=bf, labelsep=period, justification=raggedright}

\begin{center}
{\Large\textbf{Supplementary Materials}}
\end{center}

\noindent This document provides supplementary methods, tables, and figures accompanying the main manuscript. Tables and figures are numbered with an ``S'' prefix (e.g., Table~S1, Fig.~S1) to distinguish them from those in the main text.

% \tableofcontents  % optional: uncomment for an SM-only outline; note this
                     % will also pick up main-text \section commands unless
                     % you use a package like etoc for a scoped mini-TOC

%% Body content of supplementary.tex, trimmed for inclusion inside main.tex.
%% Removed vs. the standalone supplementary.tex: \documentclass/preamble,
%% \title, \begin{document}, \maketitle, the "This document provides..."
%% intro paragraph, and \tableofcontents — all of that is already handled
%% in main.tex's merged-SM block. Everything else below is unchanged.
%%
%% NOTE: this file ends with its own \bibliography{reference} call,
%% reproducing the original two-reference-list structure (one after the
%% main text, one after the SM) — both lists share the single
%% \bibliographystyle{unsrt} set once in main.tex, since bibtex only
%% allows one \bibliographystyle per document. If you'd rather have a
%% single consolidated reference list instead, delete the \bibliography
%% line at the very end of this file — the \bibliography{reference} call
%% already placed after the main sections in main.tex will still pick up
%% every \cite{} used anywhere in the SM, since bibtex indexes the whole
%% .aux file regardless of where \bibliography appears.

\section{Full Results: Main Text Tables}
\label{si:full-results}

The tables in this section are results referenced in the main manuscript's Results section.

\begin{table}[H]
	\centering
    \footnotesize
	\caption{\textbf{Treatment Effect Estimates for Primary Outcomes.}}
	\label{tab:primary}
    \scalebox{0.85}{
        \begin{tabular}{lllrrrlrr}
\toprule
Outcome & Comparison & Estimand & N & Estimate & SE & 95\% CI & p & p adjusted \\
\midrule
Trust Scale & AI Mode vs. Current & ITT & 613 & -0.337 & 0.068 & [-0.471, -0.203] & $<$0.001 & $<$0.001 \\
Trust Scale & AI Mode vs. Current & LATE & 613 & -0.359 & 0.073 & [-0.502, -0.215] & $<$0.001 & $<$0.001 \\
Trust Scale & No AI vs. Current & ITT & 647 & 0.070 & 0.056 & [-0.040, 0.180] & 0.211 & 0.145 \\
Trust Scale & No AI vs. Current & LATE & 541 & 0.144 & 0.122 & [-0.097, 0.384] & 0.241 & 0.107 \\
Click-Through Rate & AI Mode vs. Current & ITT & 714 & -0.188 & 0.018 & [-0.222, -0.153] & $<$0.001 & $<$0.001 \\
Click-Through Rate & AI Mode vs. Current & LATE & 714 & -0.199 & 0.019 & [-0.236, -0.163] & $<$0.001 & $<$0.001 \\
Click-Through Rate & No AI vs. Current & ITT & 742 & 0.027 & 0.018 & [-0.007, 0.062] & 0.124 & 0.103 \\
Click-Through Rate & No AI vs. Current & LATE & 615 & 0.088 & 0.033 & [0.023, 0.153] & 0.008 & 0.006 \\
Sessions per Day & AI Mode vs. Current & ITT & 714 & -0.920 & 0.192 & [-1.296, -0.545] & $<$0.001 & $<$0.001 \\
Sessions per Day & AI Mode vs. Current & LATE & 714 & -0.978 & 0.204 & [-1.379, -0.577] & $<$0.001 & $<$0.001 \\
Sessions per Day & No AI vs. Current & ITT & 742 & 0.100 & 0.197 & [-0.287, 0.487] & 0.612 & 0.257 \\
Sessions per Day & No AI vs. Current & LATE & 615 & 0.345 & 0.453 & [-0.545, 1.236] & 0.446 & 0.175 \\
\bottomrule
\end{tabular}

        }
\end{table}

\begin{table}[H]
    \centering
    \footnotesize
	\caption{Treatment Effect Estimates for Secondary Click Outcomes.}
	\label{tab:secondary_clicks}
    \scalebox{0.85}{
        \begin{tabular}{lllrrrlrr}
\toprule
Outcome & Comparison & Estimand & N & Estimate & SE & 95\% CI & p & p adjusted \\
\midrule
Clicks per Day & AI Mode vs. Current & ITT & 714 & -2.273 & 0.257 & [-2.777, -1.770] & $<$0.001 & $<$0.001 \\
Clicks per Day & AI Mode vs. Current & LATE & 714 & -2.415 & 0.275 & [-2.954, -1.876] & $<$0.001 & $<$0.001 \\
Clicks per Day & No AI vs. Current & ITT & 742 & 0.386 & 0.277 & [-0.157, 0.929] & 0.163 & 0.100 \\
Clicks per Day & No AI vs. Current & LATE & 615 & 1.072 & 0.641 & [-0.186, 2.330] & 0.095 & 0.044 \\
Fraction of News Clickers & AI Mode vs. Current & ITT & 714 & -0.125 & 0.032 & [-0.187, -0.063] & $<$0.001 & $<$0.001 \\
Fraction of News Clickers & AI Mode vs. Current & LATE & 714 & -0.133 & 0.034 & [-0.199, -0.067] & $<$0.001 & $<$0.001 \\
Fraction of News Clickers & No AI vs. Current & ITT & 742 & 0.071 & 0.033 & [0.006, 0.137] & 0.033 & 0.029 \\
Fraction of News Clickers & No AI vs. Current & LATE & 615 & 0.202 & 0.073 & [0.058, 0.346] & 0.006 & 0.005 \\
Fraction of Reddit Clickers & AI Mode vs. Current & ITT & 714 & -0.212 & 0.032 & [-0.275, -0.148] & $<$0.001 & $<$0.001 \\
Fraction of Reddit Clickers & AI Mode vs. Current & LATE & 714 & -0.225 & 0.035 & [-0.293, -0.157] & $<$0.001 & $<$0.001 \\
Fraction of Reddit Clickers & No AI vs. Current & ITT & 742 & -0.016 & 0.033 & [-0.080, 0.049] & 0.633 & 0.267 \\
Fraction of Reddit Clickers & No AI vs. Current & LATE & 615 & -0.049 & 0.072 & [-0.190, 0.092] & 0.495 & 0.145 \\
Fraction of Wikipedia Clickers & AI Mode vs. Current & ITT & 714 & -0.099 & 0.025 & [-0.147, -0.050] & $<$0.001 & $<$0.001 \\
Fraction of Wikipedia Clickers & AI Mode vs. Current & LATE & 714 & -0.105 & 0.027 & [-0.157, -0.053] & $<$0.001 & $<$0.001 \\
Fraction of Wikipedia Clickers & No AI vs. Current & ITT & 742 & 0.024 & 0.027 & [-0.029, 0.076] & 0.376 & 0.185 \\
Fraction of Wikipedia Clickers & No AI vs. Current & LATE & 615 & 0.059 & 0.062 & [-0.062, 0.180] & 0.336 & 0.139 \\
Fraction of Ad Clickers & AI Mode vs. Current & ITT & 714 & -0.427 & 0.026 & [-0.477, -0.377] & $<$0.001 & $<$0.001 \\
Fraction of Ad Clickers & AI Mode vs. Current & LATE & 714 & -0.454 & 0.027 & [-0.507, -0.400] & $<$0.001 & $<$0.001 \\
Fraction of Ad Clickers & No AI vs. Current & ITT & 742 & 0.024 & 0.033 & [-0.040, 0.088] & 0.470 & 0.206 \\
Fraction of Ad Clickers & No AI vs. Current & LATE & 615 & 0.098 & 0.071 & [-0.041, 0.237] & 0.166 & 0.071 \\
\bottomrule
\end{tabular}

    }
\end{table}

\begin{table}[H]
    \centering
    \footnotesize
	\caption{Treatment Effect Estimates for Secondary Search Outcomes.}
	\label{tab:secondary_search}
    \scalebox{0.85}{
        \begin{tabular}{lllrrrlrr}
\toprule
Outcome & Comparison & Estimand & N & Estimate & SE & 95\% CI & p & p adjusted \\
\midrule
Searches per Session & AI Mode vs. Current & ITT & 714 & 0.138 & 0.076 & [-0.011, 0.286] & 0.070 & 0.096 \\
Searches per Session & AI Mode vs. Current & LATE & 714 & 0.146 & 0.080 & [-0.012, 0.304] & 0.070 & 0.095 \\
Searches per Session & No AI vs. Current & ITT & 742 & -0.077 & 0.064 & [-0.203, 0.049] & 0.229 & 0.224 \\
Searches per Session & No AI vs. Current & LATE & 615 & -0.198 & 0.136 & [-0.464, 0.069] & 0.146 & 0.162 \\
Minutes per Session & AI Mode vs. Current & ITT & 714 & 0.429 & 0.116 & [0.201, 0.657] & $<$0.001 & 0.001 \\
Minutes per Session & AI Mode vs. Current & LATE & 714 & 0.456 & 0.123 & [0.214, 0.697] & $<$0.001 & 0.001 \\
Minutes per Session & No AI vs. Current & ITT & 742 & -0.217 & 0.098 & [-0.409, -0.025] & 0.027 & 0.057 \\
Minutes per Session & No AI vs. Current & LATE & 615 & -0.593 & 0.212 & [-1.009, -0.176] & 0.005 & 0.011 \\
Fraction of Question Searches & AI Mode vs. Current & ITT & 714 & 0.004 & 0.009 & [-0.014, 0.022] & 0.667 & 0.578 \\
Fraction of Question Searches & AI Mode vs. Current & LATE & 714 & 0.004 & 0.010 & [-0.015, 0.023] & 0.667 & 0.333 \\
Fraction of Question Searches & No AI vs. Current & ITT & 742 & 0.000 & 0.009 & [-0.018, 0.019] & 0.990 & 0.591 \\
Fraction of Question Searches & No AI vs. Current & LATE & 615 & 0.012 & 0.021 & [-0.028, 0.052] & 0.561 & 0.317 \\
\bottomrule
\end{tabular}

    }
\end{table}

\begin{table}[H]
    \centering
    \footnotesize
	\caption{Treatment Effect Estimates for Secondary Substitution Outcomes.}
	\label{tab:secondary_substitution}
    \scalebox{0.85}{
        \begin{tabular}{lllrrrlrr}
\toprule
Outcome & Comparison & Estimand & N & Estimate & SE & 95\% CI & p & p adjusted \\
\midrule
Fraction of Bing/DDG/Yahoo Users & AI Mode vs. Current & ITT & 714 & 0.112 & 0.024 & [0.064, 0.160] & $<$0.001 & $<$0.001 \\
Fraction of Bing/DDG/Yahoo Users & AI Mode vs. Current & LATE & 714 & 0.119 & 0.026 & [0.068, 0.170] & $<$0.001 & $<$0.001 \\
Fraction of Bing/DDG/Yahoo Users & No AI vs. Current & ITT & 742 & 0.017 & 0.019 & [-0.020, 0.055] & 0.357 & 0.312 \\
Fraction of Bing/DDG/Yahoo Users & No AI vs. Current & LATE & 615 & 0.008 & 0.039 & [-0.070, 0.085] & 0.843 & 0.728 \\
Intention to Switch to Bing & AI Mode vs. Current & ITT & 613 & 1.254 & 0.149 & [0.962, 1.546] & $<$0.001 & $<$0.001 \\
Intention to Switch to Bing & AI Mode vs. Current & LATE & 613 & 1.333 & 0.159 & [1.022, 1.645] & $<$0.001 & $<$0.001 \\
Intention to Switch to Bing & No AI vs. Current & ITT & 647 & 0.015 & 0.115 & [-0.211, 0.240] & 0.897 & 0.814 \\
Intention to Switch to Bing & No AI vs. Current & LATE & 541 & 0.098 & 0.247 & [-0.388, 0.583] & 0.692 & 0.728 \\
\bottomrule
\end{tabular}

    }
\end{table}

\begin{table}[H]
    \centering
    \footnotesize
	\caption{Treatment Effect Estimates for Secondary Survey Outcomes. Effects are presented in standardized units.}
	\label{tab:secondary_survey}
    \scalebox{0.85}{
        \begin{tabular}{lllrrrlrr}
\toprule
Outcome & Comparison & Estimand & N & Estimate & SE & 95\% CI & p & p adjusted \\
\midrule
Usefulness Scale & AI Mode vs. Current & ITT & 613 & -0.587 & 0.076 & [-0.736, -0.438] & $<$0.001 & $<$0.001 \\
Usefulness Scale & AI Mode vs. Current & LATE & 613 & -0.624 & 0.081 & [-0.783, -0.464] & $<$0.001 & $<$0.001 \\
Usefulness Scale & No AI vs. Current & ITT & 647 & 0.065 & 0.050 & [-0.033, 0.164] & 0.191 & 0.230 \\
Usefulness Scale & No AI vs. Current & LATE & 541 & 0.141 & 0.110 & [-0.076, 0.357] & 0.203 & 0.184 \\
Satisfaction Scale & AI Mode vs. Current & ITT & 613 & -0.733 & 0.073 & [-0.877, -0.589] & $<$0.001 & $<$0.001 \\
Satisfaction Scale & AI Mode vs. Current & LATE & 613 & -0.779 & 0.079 & [-0.933, -0.624] & $<$0.001 & $<$0.001 \\
Satisfaction Scale & No AI vs. Current & ITT & 647 & 0.013 & 0.049 & [-0.083, 0.110] & 0.789 & 0.487 \\
Satisfaction Scale & No AI vs. Current & LATE & 541 & 0.010 & 0.110 & [-0.206, 0.227] & 0.926 & 0.419 \\
Agency & AI Mode vs. Current & ITT & 613 & -0.657 & 0.078 & [-0.809, -0.505] & $<$0.001 & $<$0.001 \\
Agency & AI Mode vs. Current & LATE & 613 & -0.699 & 0.082 & [-0.860, -0.537] & $<$0.001 & $<$0.001 \\
Agency & No AI vs. Current & ITT & 647 & 0.095 & 0.059 & [-0.021, 0.211] & 0.108 & 0.161 \\
Agency & No AI vs. Current & LATE & 541 & 0.188 & 0.128 & [-0.063, 0.439] & 0.141 & 0.165 \\
Personalization/Relevance & AI Mode vs. Current & ITT & 613 & -0.423 & 0.078 & [-0.576, -0.270] & $<$0.001 & $<$0.001 \\
Personalization/Relevance & AI Mode vs. Current & LATE & 613 & -0.450 & 0.083 & [-0.613, -0.287] & $<$0.001 & $<$0.001 \\
Personalization/Relevance & No AI vs. Current & ITT & 647 & 0.066 & 0.060 & [-0.051, 0.184] & 0.269 & 0.243 \\
Personalization/Relevance & No AI vs. Current & LATE & 541 & 0.229 & 0.130 & [-0.026, 0.484] & 0.078 & 0.111 \\
Preference for AI Mode (News) & AI Mode vs. Current & ITT & 1836 & -0.123 & 0.066 & [-0.252, 0.006] & 0.061 & 0.102 \\
Preference for AI Mode (News) & AI Mode vs. Current & LATE & 1836 & -0.131 & 0.070 & [-0.268, 0.006] & 0.060 & 0.099 \\
Preference for AI Mode (News) & No AI vs. Current & ITT & 1938 & -0.065 & 0.063 & [-0.189, 0.058] & 0.301 & 0.251 \\
Preference for AI Mode (News) & No AI vs. Current & LATE & 1623 & -0.158 & 0.139 & [-0.430, 0.113] & 0.253 & 0.184 \\
Trust in AI Mode (News) & AI Mode vs. Current & ITT & 1836 & -0.078 & 0.067 & [-0.209, 0.053] & 0.243 & 0.241 \\
Trust in AI Mode (News) & AI Mode vs. Current & LATE & 1836 & -0.083 & 0.071 & [-0.222, 0.056] & 0.241 & 0.184 \\
Trust in AI Mode (News) & No AI vs. Current & ITT & 1938 & -0.061 & 0.065 & [-0.188, 0.067] & 0.351 & 0.275 \\
Trust in AI Mode (News) & No AI vs. Current & LATE & 1623 & -0.151 & 0.142 & [-0.429, 0.128] & 0.289 & 0.184 \\
\bottomrule
\end{tabular}

        }
\end{table}
\section{Materials and Methods}
\label{si:methods}
\subsection{Participant Recruitment}
We recruited participants from two channels: Prolific and Northeastern students enrolled in a work-study program with one of the authors. Participants were recruited in waves between March 17 and March 19, 2026. We limited recruitment to US-based participants, and screened for individuals who (1) were 18 years or older, (2) used Google Chrome as their primary browser, and (3) used Google Search as their primary search engine. After giving informed consent, participants completed a pre-experiment survey which elicited their familiarity with and sentiment towards LLM applications, as well as perceived trust, usefulness, satisfaction, agency, and personalization/relevance of information-seeking on Google. Participants then installed our browser extension. Prolific participants were compensated \$2 for completing the screening, pre-experiment survey and installing the browser extension, and \$8 for finishing the study. Northeastern students were paid \$17 per hour through their work-study program.
\subsection{Interventions}
\label{si:interventions}
Our study implemented three search conditions via a Chromium browser extension:
\begin{enumerate}
    \item \textbf{No AI Search}: Hides Google's AI Overviews, including those at the top of the results page, in the middle of the results page, and nested within People Also Ask components. AI Mode searches are redirected to general search at the network level. The ``AI Mode'' tab on the Google Search page is hidden. The ``AI Mode'' prompt on the Google home page is hidden. Unfortunately, browser extensions cannot hide the ``AI Mode'' prompt in the Chrome browser’s address bar.
    \item \textbf{Current Search}: No interventions are applied, so this is the default experience of Google Search. AI Overviews may be present, AI Mode is accessible, and ``AI Mode'' tabs/prompts are visible.
    \item \textbf{AI Mode Search}: All Google searches are redirected to AI Mode at the network level. The ``All'' and ``Web'' tabs on the Google search page are hidden, which direct to general search and web results, respectively. Search verticals (e.g., news, shopping, images) are still accessible.
\end{enumerate}
\subsection{Experimental Platform}
\label{si:platform}
We developed a custom experimental platform that enabled us to collect participants' web browsing behavior, deploy custom surveys, and make in-situ changes to their Google Search experience.
\paragraph{Browser extension} The browser extension handles enforcing the three Google Search conditions and logging browsing activity, HTML snapshots of no AI, Current and AI Mode results, and click events that originate from Google Search, Gemini and ChatGPT. The extension is compatible with all Chromium-derived browsers, including Google Chrome, and is published on the Chrome Web Store (\url{https://chromewebstore.google.com/detail/cedfhkdnooeicfiaheamppdckbidiaah}).
The extension manages enforcing the conditions by injecting a content script into Google Search pages, which detected and modified the rendered DOM in real time. Participants were randomly assigned to one of three conditions managed through a field stored in the extension's local storage. Upon each Google Search page load, the extension retrieved this assignment and applied the corresponding intervention. In the No AI Search condition, the extension continuously monitored DOM mutations via a MutationObserver. When an AI Overview container was detected, its CSS `display' property was set to `none'. The `AI Mode' search tab was similarly hidden to prevent participants from navigating to it. In the AI Mode condition, the extension suppressed the default `All' and `Web' search tabs, and redirected all Google searches to AI Mode at the network level (i.e., URL parameter udm=50). Participants in this condition received a one-time disclosure modal on their first exposure to AI Mode, informing them that they would be using AI-driven search for the study duration. The Current Search condition received no DOM modifications.
The extension passively recorded participants' interactions across Google Search, Google AI Mode, Google Gemini and ChatGPT. On Google Search pages, the extension captured the full HTML of the results page, along with the AI Overview container. On AI Mode, Gemini and ChatGPT, a MutationObserver tracked each AI response, capturing HTML and the sequential conversational turn order. Across all platforms, outbound link clicks were logged.
\paragraph{Web application} The study's web application was built with Django and deployed on a server hosted by Northeastern University, backed by a PostgreSQL database. The application managed participant onboarding, accounts, condition assignment, survey administration, and data collection from the browser extension.
Upon enrollment, participants were directed to the web application and guided through an onboarding flow: providing informed consent, authenticating via Google social auth using their Prolific or other email address, completing a demographic survey and pre-survey, installing the browser extension from the Chrome Web Store, and verifying the extension was active before beginning the study.
During the study period, the extension communicated with the web application via dedicated endpoints to store scraped data, retrieve condition assignments, and manage authenticated sessions. All communication occurred over HTTPS, with scraped HTML, click events, and URLs stored in the database for later analysis.
At the conclusion of the study period, participants were directed to the web application via a URL sent to their Prolific inbox or email to complete the post-survey.
\subsection{Outcome Measures}
\label{si:outcomes}
\subsubsection*{Primary Outcome Measures}
\begin{itemize}
    \item \textit{Trust.} We measure trust in Google responses using a validated 5-item scale~\cite{flanagin2000perceptions} that has been used in recent work evaluating AI search~\cite{li_human_2025}. The five items are: ``Over the last 7 days, how [believable / trustworthy / accurate / biased / complete] was the information you found searching on Google?'' Each item is measured on a 7-point Likert scale (\textit{not at all} to \textit{extremely}). The bias item is reverse-coded.

    \item \textit{Click-Through Rate.} The total number of external clicks originating from Google divided by the total number of Google searches. The numerator excludes clicks to google.com (e.g., follow-up searches). The denominator excludes searches on Google's vertical search pages (e.g., images, news, shopping). Each turn of a conversation in AI Mode is counted as a separate search. This is a user-level metric.

    \item \textit{Search Sessions.} Searches on Google's vertical search pages are excluded. Searches are grouped into sessions such that if the inter-arrival time between two searches exceeds 5 minutes, the second query is considered a new session. When evaluated against 100 randomly sampled, manually annotated query pairs from the pilot, this method produced an F1 score of 0.84.
\end{itemize}
\begin{figure}
    \centering
    \includegraphics[width=\linewidth]{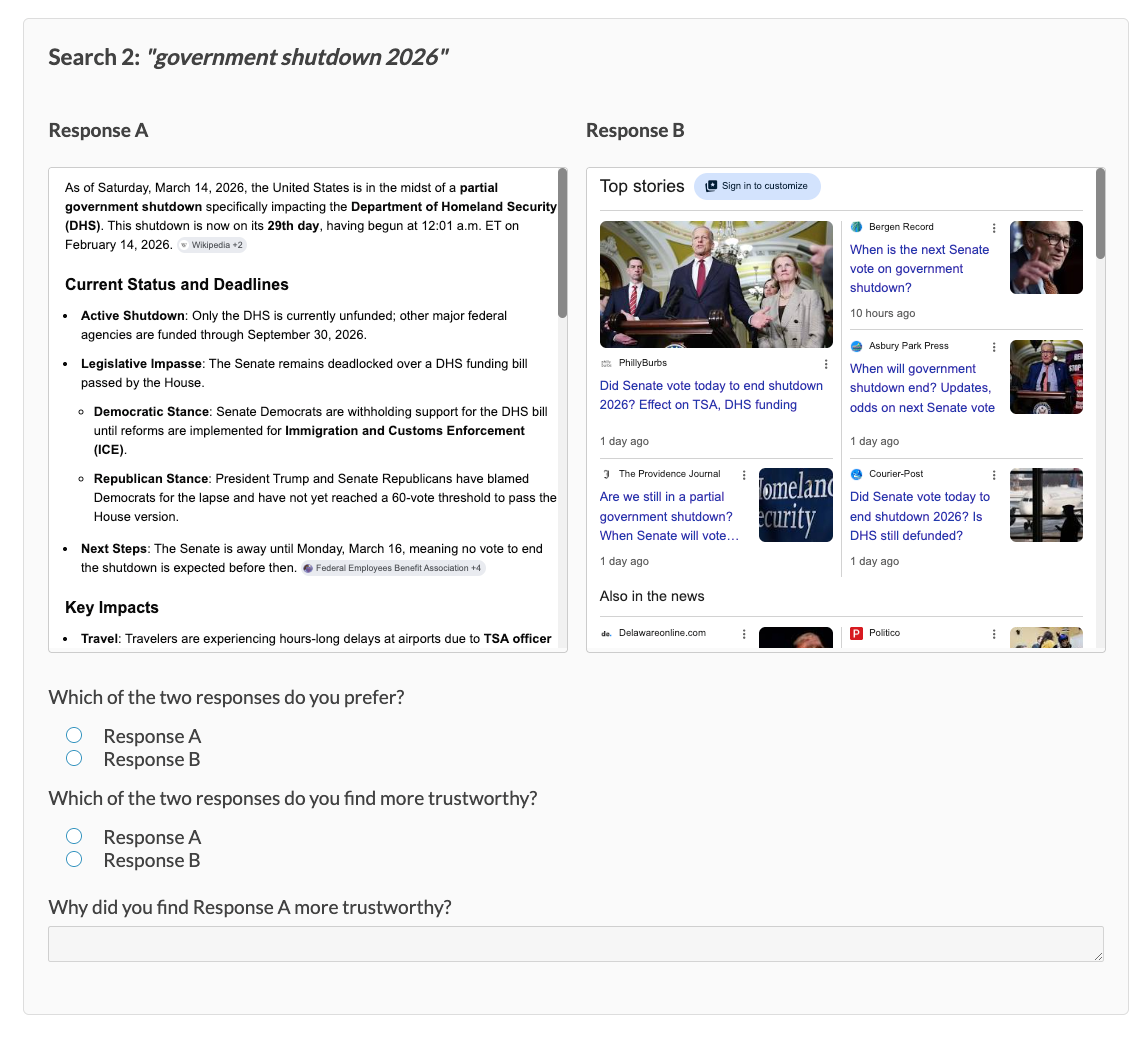}
    \caption{Head to head survey for a news-related query, prompting users to answer questions comparing Response A (AI Mode result for news query) to Response B (Current search result for news query).}
    \label{fig:head_to_head_news}
\end{figure}
\subsubsection*{Secondary Survey Outcome Measures}
\begin{itemize}
    \item \textit{Usefulness.} We measure usefulness using a validated 5-item scale~\cite{sharma_generative_2024, van1997simple}. The five items are: ``Over the last 7 days, information-seeking on Google was:'' [useless vs.\ useful; bad vs.\ good; ineffective vs.\ effective; worthless vs.\ helpful; sleep-inducing vs.\ stimulating]. Each item is measured on a 5-point scale.

    \item \textit{Satisfaction.} We measure satisfaction using a validated 4-item scale~\cite{sharma_generative_2024, van1997simple}. The four items are: ``Over the last 7 days, information-seeking on Google was:'' [unpleasant vs.\ pleasant; annoying vs.\ nice; irritating vs.\ likeable; undesirable vs.\ desirable]. Each item is measured on a 5-point scale.

    \item \textit{Agency.} Measured through a single statement: ``Over the last 7 days, I had agency over the information-seeking process on Google,'' on a 7-point Likert scale (\textit{strongly disagree} to \textit{strongly agree}).

    \item \textit{Personalization.} Measured through a single statement: ``Over the last 7 days, information-seeking on Google led to results that were personalized and relevant to me,'' on a 7-point Likert scale (\textit{strongly disagree} to \textit{strongly agree}).

    \item \textit{Preference for AI Mode (News Context).} We show participants head-to-head comparisons of responses from AI Mode Search and Current Search, asking ``Which of the two responses do you prefer?'' (Figure~\ref{fig:head_to_head_news}). Each participant is shown the results of three queries randomly sampled from 10 pre-defined trending news queries, manually selected from Google Trends the day before the final survey is released. The order of responses is randomized.

    \item \textit{Trust in AI Mode (News Context).} Using the same design as above, participants are asked ``Which of the two responses do you find more trustworthy?''

\end{itemize}
The selected trending news queries were (1) ``tsa agents government shutdown'', (2) ``iran israel war ceasefire'', (3) ``laguardia plane crash'', (4) ``us treasury insolvency'', (5) ``us army enlistment age'', (6) ``meta social media addiction trial'', (7) ``retirement benefits social security'', (8) ``no kings protest near me'', (9) ``covid 19 cicada variant'', and (10) ``ice at airports''.
\subsubsection*{Secondary Click Outcomes} All secondary click outcomes except total clicks are binary user-level variables (e.g., did user $X$ ever click through to a news site).
\begin{itemize}
    \item\textit{Clicks.} Total external clicks originating from Google, excluding clicks to [www.google.com](https://www.google.com).
    \item\textit{Clicks to news websites $> 0$.} News domains are identified using a comprehensive, actively curated list of news domains~\cite{yang2025newsdomains}.
    \item\textit{Clicks to specific domains $> 0$.} Any domain receiving at least 5\% of total clicks originating from Google during the baseline period. During the pilot, only reddit.com and wikipedia.org met this criterion.
    \item\textit{Ad clicks $> 0$.} Ad clicks are identified using the \texttt{gclid} URL parameter~\cite{google_gclid}.
\end{itemize}
\subsubsection*{Secondary Search Outcomes} All variables in this section exclude searches on Google's vertical search pages. All are user-level metrics.
\begin{itemize}
    \item\textit{Searches per session.} Total searches divided by total sessions. Each turn of a conversation in AI Mode is counted as a separate search.
    \item\textit{Minutes per session.} Total active minutes on Google Search pages divided by total sessions. Time is only accumulated when a Google Search page is in the browser's foreground.
    \item\textit{Fraction of question searches.} The fraction of searches beginning with ``who'', ``what'', ``where'', ``why'', ``when'', or ``how''.
\end{itemize}
\subsubsection*{Secondary Substitution Outcomes}
\begin{itemize}
    \item\textit{Competing search engine use $> 0$.} A binary user-level variable indicating whether the user ran any searches on Bing, Yahoo, or DuckDuckGo.
    \item\textit{Intention to switch to Bing.} Measured by asking ``If your search experience from the last 7 days remained permanent, how likely would you be to switch to Bing?'' on a 7-point Likert scale (\textit{extremely unlikely} to \textit{extremely likely}).
\end{itemize}
\subsection{Hypotheses}
\label{si:hypotheses}
\subsubsection*{Primary Hypotheses}
\paragraph{H1 (Trust):} Based on existing work on trust in AI search responses~\cite{li_human_2025}.
\begin{itemize}
    \item[H1a:] AI Mode Search will decrease overall trust in responses compared to Current Search.
    \item[H1b:] No AI Search will increase overall trust in responses compared to Current Search.
\end{itemize}
\paragraph{H2 (Click-Through Rate):} Based on reports from publishers~\cite{simonetti_news_2025} and observational data~\cite{chapekis2025google}.
\begin{itemize}
    \item[H2a:] AI Mode Search will decrease click-through rate to external sites compared to Current Search.
    \item[H2b:] No AI Search will increase click-through rate to external sites compared to Current Search.
\end{itemize}
\paragraph{H3 (Search Sessions):} Based on Google's claims about the impact of AI features~\cite{reid_ai_2025}.
\begin{itemize}
    \item[H3a:] AI Mode Search will increase the number of search sessions compared to Current Search.
    \item[H3b:] No AI Search will decrease the number of search sessions compared to Current Search.
\end{itemize}
\subsubsection*{Secondary Hypotheses: Survey Outcomes}
\paragraph{H4 (Satisfaction, Usefulness, Agency):} We don’t have a strong hypothesis (H4a). AI Mode Search could decrease [satisfaction with, usefulness from, agency over] responses compared to Current Search because of information quality concerns. It could also increase these outcomes because of perceived efficiency, personalization, and/or relevance. Analogous reasoning for No AI Search (H4b).
\paragraph{H4 (Personalization/Relevance):} Based on Google's claims~\cite{google_personal_intelligence_search}.
\begin{itemize}
    \item[H4c:] AI Mode Search will increase perceptions of responses being personalized and relevant.
    \item[H4d:] No AI Search will decrease perceptions of responses being personalized and relevant.
\end{itemize}
\paragraph{H4 (News Context):}
\begin{itemize}
    \item[H4e:] AI Mode Search will decrease preference for and trust in AI Mode in the context of news search.
    \item[H4f:] No AI Search will have no impact on these outcomes.
\end{itemize}
\subsubsection*{Secondary Hypotheses: Click Outcomes}
\paragraph{H5 (Clicks to domains):}
\begin{itemize}
    \item[H5a:] AI Mode Search will decrease total clicks, clicks to news sites, clicks to specific domains, and ad clicks compared to Current Search.
    \item[H5b:] No AI Search will increase these outcomes compared to Current Search.
\end{itemize}
\subsubsection*{Secondary Hypotheses: Search Behavior}
\paragraph{H6 (Session Characteristics):}
\begin{itemize}
    \item[H6a:] AI Mode Search will increase minutes per session and the fraction of question-form searches compared to Current Search.
    \item[H6b:] No AI Search will decrease these outcomes compared to Current Search.
\end{itemize}
\paragraph{H6 (Searches per Session):} No strong hypothesis; AI Mode Search could increase or decrease searches per session depending on whether it encourages follow-up queries or resolves them preemptively (H6c), with analogous reasoning for No AI Search (H6d).
\subsubsection*{Secondary Hypotheses: Search Engine Substitution}
\paragraph{H7 (Substitution):}
\begin{itemize}
    \item[H7a:] AI Mode Search will increase substitution to other search engines (Bing, DuckDuckGo, Yahoo) compared to Current Search.
    \item[H7b:] No AI Search will have no impact on substitution to other search engines.
\end{itemize}
\section{Participant Demographics}
\label{si:demographics}
\begin{figure}[H]
	\centering
    \includegraphics[width=\textwidth]{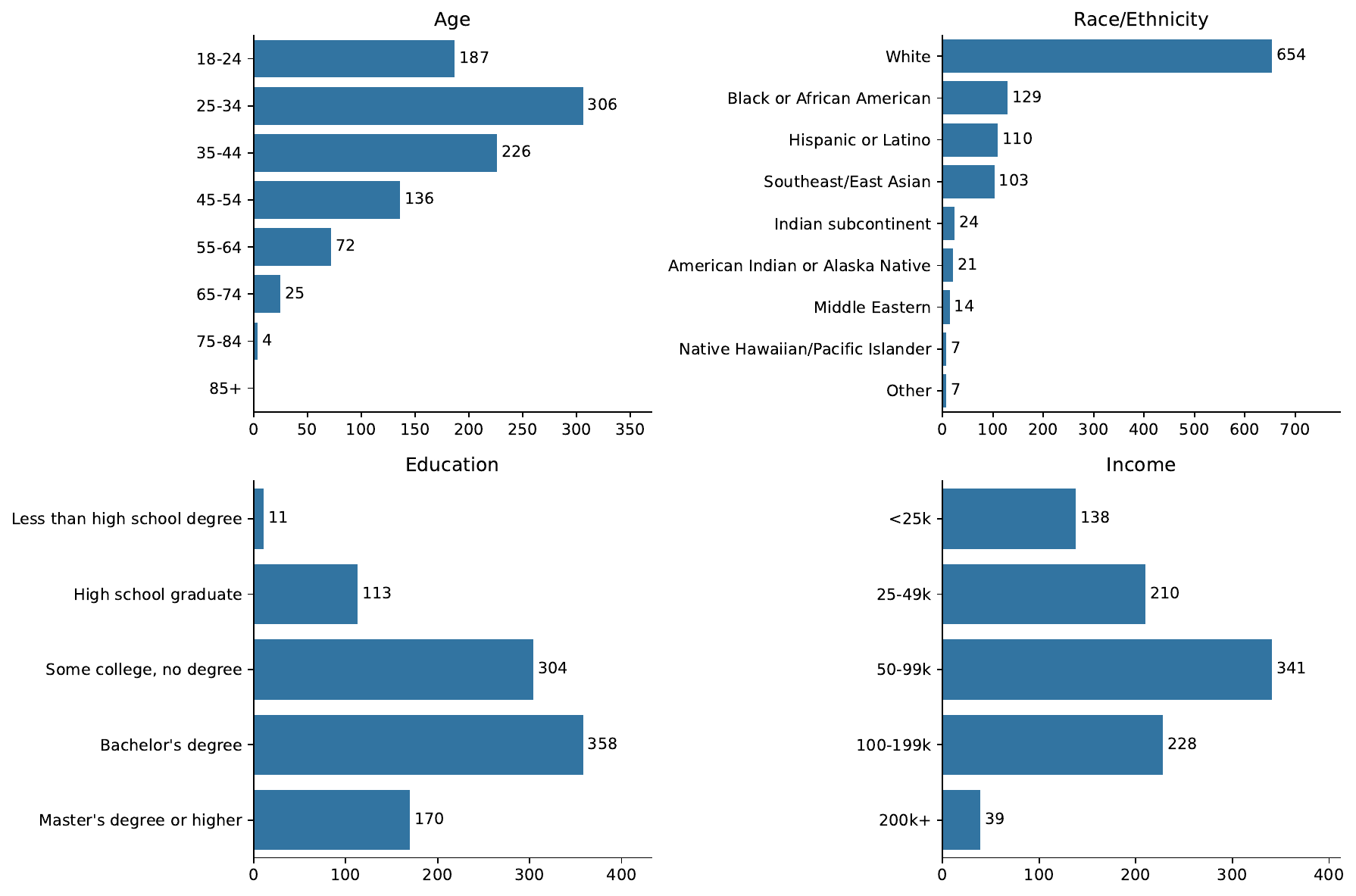}
    % Captions go below figures
    \caption{\textbf{Distribution of demographics among participants who completed the post-survey.}}
    \label{fig:participant_demographics} % give each figure a logical label name
\end{figure}

Of the 956 participants who completed the final survey, 50\% identified as women, 48\% as men, and 2\% as another gender. Politically, participants leaned Democratic, as 59\% identified as Democrat (30\% strong, 13\% weak, 16\% leaning), 21\% as Independent, and 20\% as Republican (10\% leaning, 4\% weak, 6\% strong). Figure~\ref{fig:participant_demographics} presents the demographic composition of the survey analysis sample across four other characteristics: age, race, education, and income. Table~\ref{tab:demographics} summarizes the demographic composition of the enrolled sample ($N = 1,444$), the behavioral analysis sample ($N = 1,100$) and the survey analysis sample ($N = 956$).
\begin{table}[H] % Do not use \begin{table*}
	\centering
    \footnotesize
	% Captions go above tables
	\caption{\textbf{Participant Demographics.} Demographic composition of the enrolled sample, the behavioral analysis sample, and the survey analysis sample.}
	\label{tab:demographics}
    \scalebox{0.85}{
        \begin{tabular}{lrrr}
\toprule
 & Enrolled (N=1444) & Behavioral (N=1100) & Survey (N=956) \\
\midrule
Age: 18-24 (\%) & 0.222 & 0.212 & 0.196 \\
Age: 25-34 (\%) & 0.298 & 0.309 & 0.320 \\
Age: 35-44 (\%) & 0.235 & 0.231 & 0.236 \\
Age: 45-54 (\%) & 0.142 & 0.145 & 0.142 \\
Age: 55-64 (\%) & 0.075 & 0.075 & 0.075 \\
Age: 65-74 (\%) & 0.024 & 0.025 & 0.026 \\
Age: 75-84 (\%) & 0.005 & 0.005 & 0.004 \\
Age: 85+ (\%) & 0.000 & 0.000 & 0.000 \\
Education: Less than high school degree (\%) & 0.013 & 0.010 & 0.012 \\
Education: High school graduate (\%) & 0.126 & 0.120 & 0.118 \\
Education: Some college, no degree (\%) & 0.329 & 0.335 & 0.318 \\
Education: Bachelor's degree (\%) & 0.357 & 0.357 & 0.374 \\
Education: Master's degree or higher (\%) & 0.175 & 0.178 & 0.178 \\
Income: $<$25k (\%) & 0.151 & 0.145 & 0.144 \\
Income: 25-49k (\%) & 0.219 & 0.219 & 0.220 \\
Income: 50-99k (\%) & 0.350 & 0.359 & 0.357 \\
Income: 100-199k (\%) & 0.228 & 0.230 & 0.238 \\
Income: 200k+ (\%) & 0.053 & 0.046 & 0.041 \\
Political Party: Strong Democrat (\%) & 0.278 & 0.292 & 0.298 \\
Political Party: Weak Democrat (\%) & 0.115 & 0.124 & 0.132 \\
Political Party: Lean Democrat (\%) & 0.166 & 0.165 & 0.159 \\
Political Party: Independent (\%) & 0.234 & 0.222 & 0.214 \\
Political Party: Lean Republican (\%) & 0.098 & 0.100 & 0.099 \\
Political Party: Weak Republican (\%) & 0.043 & 0.039 & 0.038 \\
Political Party: Strong Republican (\%) & 0.066 & 0.058 & 0.060 \\
Gender: Woman (\%) & 0.513 & 0.499 & 0.500 \\
Gender: Man (\%) & 0.467 & 0.480 & 0.477 \\
Gender: Nonbinary (\%) & 0.018 & 0.020 & 0.022 \\
Race/Ethnicity: White (\%) & 0.669 & 0.679 & 0.684 \\
Race/Ethnicity: Black or African American (\%) & 0.159 & 0.145 & 0.135 \\
Race/Ethnicity: Hispanic or Latino (\%) & 0.107 & 0.109 & 0.115 \\
Race/Ethnicity: Southeast/East Asian (\%) & 0.103 & 0.106 & 0.108 \\
Race/Ethnicity: Indian subcontinent (\%) & 0.022 & 0.025 & 0.025 \\
Race/Ethnicity: American Indian or Alaska Native (\%) & 0.023 & 0.024 & 0.022 \\
Race/Ethnicity: Middle Eastern (\%) & 0.017 & 0.017 & 0.015 \\
Race/Ethnicity: Native Hawaiian/Pacific Islander (\%) & 0.006 & 0.006 & 0.007 \\
Race/Ethnicity: Other (\%) & 0.006 & 0.006 & 0.007 \\
\bottomrule
\end{tabular}

    }
\end{table}
\section{Analysis Plan}
\label{si:analysis}
\subsection{Randomization} Participants were randomly assigned to one of three experimental conditions at the time of extension installation, whereby each participant was independently assigned with equal probability and no constraints imposed on group balance.
\subsection{Power analysis}
\label{supp:power}
To determine minimum detectable effects (MDEs) at various sample sizes, we ran a power analysis based on a pilot study with $N=72$ participants (61 from Prolific and 11 from a paid, undergraduate work-study program at Northeastern). The power analysis used power = 0.8, alpha = 0.05 / H hypothesis tests, and the residual outcome variance after adjusting for the pre-treatment outcome. H came from our pre-registered multiple testing plan for different groups of covariates (e.g., for primary outcomes, H = 3 outcomes * 2 contrasts = 6 hypothesis tests). The residual outcome variance was calculated as the unadjusted outcome variance * (1 - out-of-sample $R^2$). Out-of-sample $R^2$ was estimated using OLS regression and 10-fold cross validation. We used out-of-sample $R^2$ to estimate residual variance conservatively with a small pilot.
Our pre-registration reported MDEs corresponding to $N = 1200$ total participants. Our final analysis sample contains $N = 1100$ participants with behavioral data and $N = 956$ participants with survey data. Therefore, we report MDEs with respect to these realized sample sizes here. Note that MDEs are still calculated using pilot data, just assuming a smaller sample size. We were powered to detect minimum effects of 0.24 (4\%) on searcher trust, 0.07 (16\%) on click-through rate, and 0.61 (16\%) on search sessions per day.
For secondary survey outcomes, we were powered to detect minimum effects of 0.19 (5\%) on satisfaction, 0.16 (4\%) on usefulness, and 0.42 (10\%) on agency. We did not measure personalization/relevance, preference for AI Mode in the context of news search, and trust in AI Mode in the context of news search during the pilot. For secondary click outcomes, we were powered to detect minimum effects of 1.3 (35\%) on clicks per day, 0.13 (28\%) on the proportion of users clicking to Reddit, 0.12 (40\%) on the proportion of users clicking to Wikipedia, 0.13 (35\%) on the proportion of users clicking to news sites, and 0.14 (33\%) on the proportion of users clicking on ads. For secondary search outcomes, we were powered to detect minimum effects of 0.26 (12\%) on searches per session, 0.42 (33\%) on minutes per session and 0.02 (34\%) on the fraction of question searches. For secondary substitution outcomes, we were powered to detect a minimum effect of 0.08 (79\%) on the proportion of people using a competing search engine. We did not measure long-term intention to switch to Bing during the pilot.
\subsection{Main Analyses}
\label{si:main-analyses}
Following our pre-registration, for all user-level outcomes, we run an OLS regression of the outcome on participants’ treatment assignment, controlling for the baseline value of the outcome and recruitment channel. We use HC2 robust standard errors.
For the news-related survey outcomes, each participant answered questions about three queries that are randomly sampled from a set of 10. Following our pre-registration, for these two outcomes, we run an OLS regression of the outcome on participants’ treatment assignment at the question-level, with fixed effects for question ID. We control for baseline trust and recruitment channel and cluster standard errors at the user-level.
Our pre-registration did not specify how we would estimate LATEs in the case of non-compliance. We run a two-stage least squares (2SLS) regression of the outcome on participants' treatment exposure (i.e., compliance rate), instrumented by treatment assignment, controlling for the baseline value of the outcome and recruitment channel. We use HC1 robust standard errors, which are supported in \texttt{Python}'s \texttt{linearmodels} package.
% \jgnote{Add comments about LATE assumptions, particularly exclusion restriction.}
Following our pre-registration, we recode any missing covariate values to the overall mean from the baseline period. Following our pre-registration, we also truncate behavioral measures (sessions per day, clicks per day, click-through rate, searches per session, and minutes per session) at the 99th percentile to deal with possible outliers. We calculate 99th percentiles using data from the baseline period. We do not truncate behavioral proportions. We also truncate the focus duration of individual searches at the 99th percentile (24.4 minutes).
\subsubsection*{Multiple Comparisons}
We implement False Discovery Rate (FDR) adjustment using the method in ~\cite{anderson2008multiple} with the following groups of variables:
\begin{itemize}
    \item For $K1=3$ primary outcomes: Sharpened FDR-adjusted p-values with $K1*2$ hypothesis tests (multiplied by 2 because 2 condition contrasts)
    \item For $K2=6$ survey-related secondary outcomes: Sharpened FDR-adjusted p-values with $(K2+1)*2$ hypothesis tests, where the extra outcome is the primary trust outcome
    \item For $K2=5$ click-related secondary outcomes: Sharpened FDR-adjusted p-values with $(K2+1)*2$ hypothesis tests, where the extra outcome is the primary CTR outcome
    \item For $K2=3$ search-related secondary outcomes: Sharpened FDR-adjusted p-values with $(K2+1)*2$ hypothesis tests, where the extra outcome is the primary sessions outcome
    \item For $K2=2$ substitution-related secondary outcomes: Sharpened FDR-adjusted p-values with $K2*2$ hypothesis tests
    \item For $L1=4$ (i.e., 2 moderators * 2 levels): Sharpened FDR-adjusted p-values with $K1*L1*2$ hypothesis tests.
\end{itemize}
For each group of variables, we adjust the set of ITT and LATE estimates separately because they represent the same group of hypothesis tests.
\section{Validity Checks}
\label{si:validity}
\subsection{Covariate balance}
\begin{table} % Do not use \begin{table*}
	\centering
    \footnotesize
	% Captions go above tables
	\caption{\textbf{Covariate Balance.} Baseline covariate means across treatment groups.}
	\label{tab:balance}
    \scalebox{0.85}{
        \begin{tabular}{lrrr}
\toprule
 & No AI & Current Search & AI Mode \\
\midrule
Sessions Per Day & 2.896 & 3.465 & 3.183 \\
Clicks Per Day & 3.130 & 3.455 & 3.332 \\
Trust & 4.432 & 4.384 & 4.425 \\
Satisfaction & 3.884 & 3.887 & 3.900 \\
Usefulness & 4.169 & 4.132 & 4.132 \\
Agency & 4.632 & 4.659 & 4.671 \\
Personalization/Relevance & 5.087 & 5.031 & 4.992 \\
LLM Familiarity & 5.195 & 5.071 & 5.150 \\
LLM Overall Sentiment & 3.772 & 3.646 & 3.698 \\
LLM Use Frequency & 3.638 & 3.512 & 3.637 \\
Race/Ethnicity: White (\%) & 0.685 & 0.644 & 0.676 \\
Race/Ethnicity: Black or African American (\%) & 0.140 & 0.169 & 0.168 \\
Race/Ethnicity: Southeast/East Asian (\%) & 0.083 & 0.127 & 0.103 \\
Race/Ethnicity: Hispanic or Latino (\%) & 0.120 & 0.094 & 0.107 \\
Race/Ethnicity: Other (\%) & 0.031 & 0.045 & 0.041 \\
Gender: Woman (\%) & 0.518 & 0.514 & 0.507 \\
Gender: Man (\%) & 0.467 & 0.463 & 0.472 \\
Gender: Nonbinary (\%) & 0.016 & 0.022 & 0.021 \\
Age: 18-24 (\%) & 0.219 & 0.207 & 0.238 \\
Age: 25-34 (\%) & 0.307 & 0.303 & 0.283 \\
Age: 35-44 (\%) & 0.240 & 0.236 & 0.230 \\
Age: 45-54 (\%) & 0.124 & 0.149 & 0.154 \\
Age: 55+ (\%) & 0.110 & 0.105 & 0.094 \\
Education: High school graduate (\%) & 0.144 & 0.143 & 0.131 \\
Education: Some college, no degree (\%) & 0.339 & 0.327 & 0.320 \\
Education: Bachelor's degree (\%) & 0.354 & 0.330 & 0.384 \\
Education: Master's degree or higher (\%) & 0.163 & 0.200 & 0.164 \\
Political Party: Democrat (\%) & 0.541 & 0.543 & 0.591 \\
Political Party: Independent (\%) & 0.258 & 0.227 & 0.216 \\
Political Party: Republican (\%) & 0.201 & 0.229 & 0.193 \\
Income: $<$25k (\%) & 0.167 & 0.167 & 0.119 \\
Income: 25-49k (\%) & 0.230 & 0.205 & 0.220 \\
Income: 50-99k (\%) & 0.352 & 0.356 & 0.341 \\
Income: 100k+ (\%) & 0.250 & 0.272 & 0.320 \\
Channel: NEU (\%) & 0.047 & 0.033 & 0.037 \\
Channel: Prolific (\%) & 0.953 & 0.967 & 0.963 \\
\bottomrule
\end{tabular}

    }
\end{table}
Following our pre-registration, we use randomization inference to test whether observed covariate imbalances are larger than would be expected by chance\cite{lin2016green}. Specifically, we fit a multinomial logistic regression of treatment assignment on baseline covariates and calculate a heteroskedasticity-robust Wald statistic for the hypothesis that all coefficients are equal to zero. We use randomization inference with 10,000 random assignments to calculate the p-value associated with the test statistic.
Following our pre-registration, we included the following covariates: recruitment channel, demographics (race/ethnicity, gender identity, age, education, political party, household income), baseline survey measures (trust, satisfaction, usefulness, agency, personalization/relevance, LLM familiarity, LLM sentiment, and LLM use frequency), and baseline outcomes (search sessions per day and clicks per day). We coarsened all demographic covariates except for gender identity.
We do not find evidence of covariate imbalance ($p$ = 0.305). Table \ref{tab:balance} shows covariate means in each treatment group.
\subsection{Compliance and manipulation checks}
\begin{figure}[ht] % Do not use \begin{figure*}
	\centering
    \includegraphics[width=0.6\textwidth]{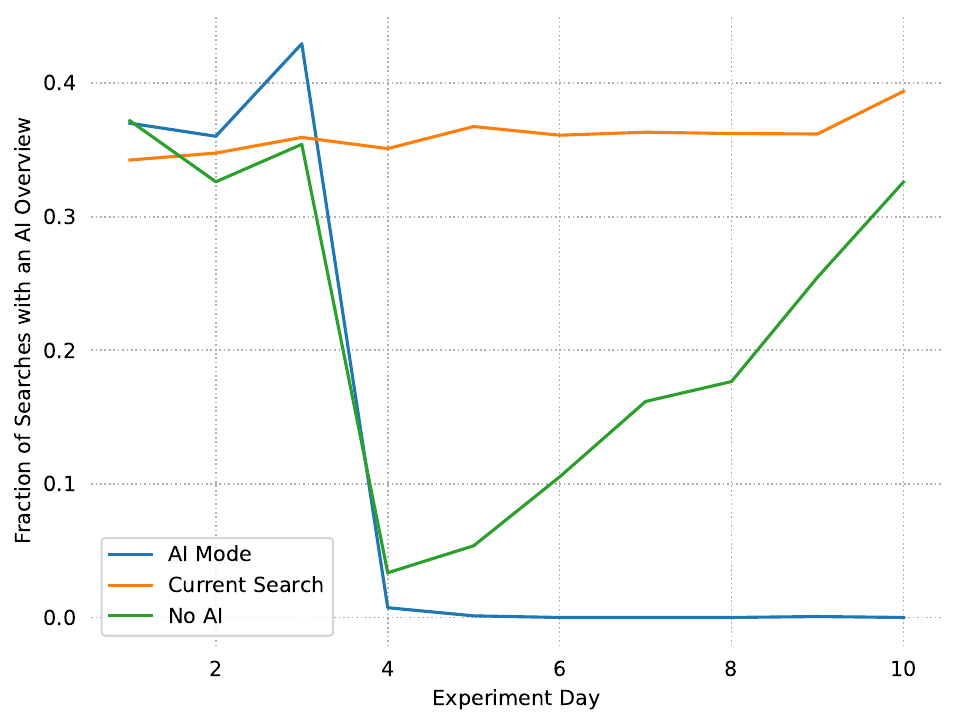}
    % Captions go below figures
    \caption{\textbf{Fraction of Searches with AI Overviews Across Experiment Days.} The fraction of searches with an AI Overview across conditions and experiment days.}
    \label{fig:ai_overview_visible_time} % give each figure a logical label name
\end{figure}
\begin{figure}[ht] % Do not use \begin{figure*}
	\centering
    \includegraphics[width=0.6\textwidth]{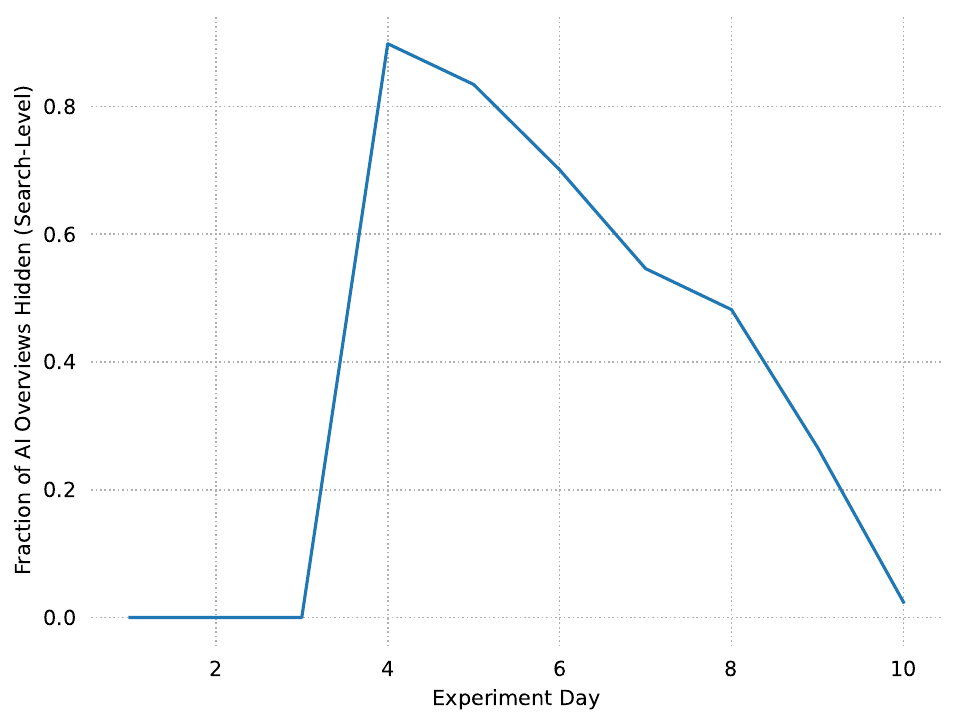}
    % Captions go below figures
    \caption{\textbf{Fraction of AI Overviews Hidden Across Experiment Days.} The fraction of AI Overviews hidden in the No AI group on each experiment day.}
    \label{fig:no_ai_compliance_time} % give each figure a logical label name
\end{figure}
\begin{figure}[ht]
    \centering
    \includegraphics[width=0.6\textwidth]{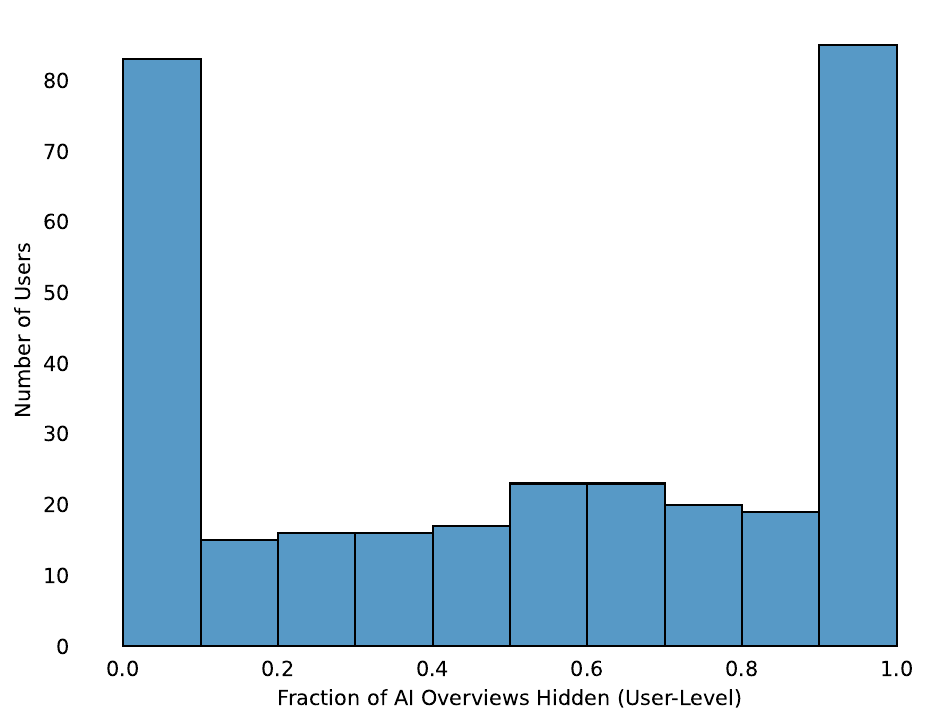}
    % Captions go below figures
    \caption{\textbf{Fraction of AI Overviews Hidden Across Participants.} The fraction of AI Overviews successfully hidden across participants in the No AI group.}
    \label{fig:no_ai_compliance_users} % give each figure a logical label name
\end{figure}
\begin{figure}[ht]
    \centering
    \includegraphics[width=0.6\textwidth]{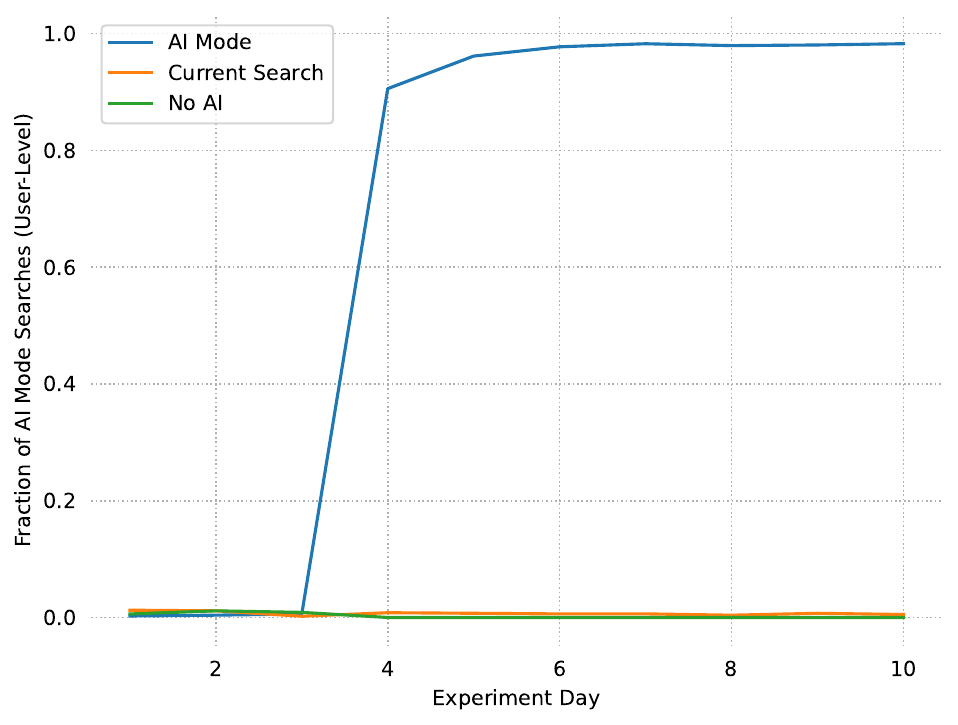}
    % Captions go below figures
    \caption{\textbf{Fraction of Searches Routed to AI Mode (User-Level).} Each user is given equal weight.}
    \label{fig:ai_mode_compliance_users} % give each figure a logical label name
\end{figure}
\begin{figure}[ht]
    \centering
    \includegraphics[width=0.6\textwidth]{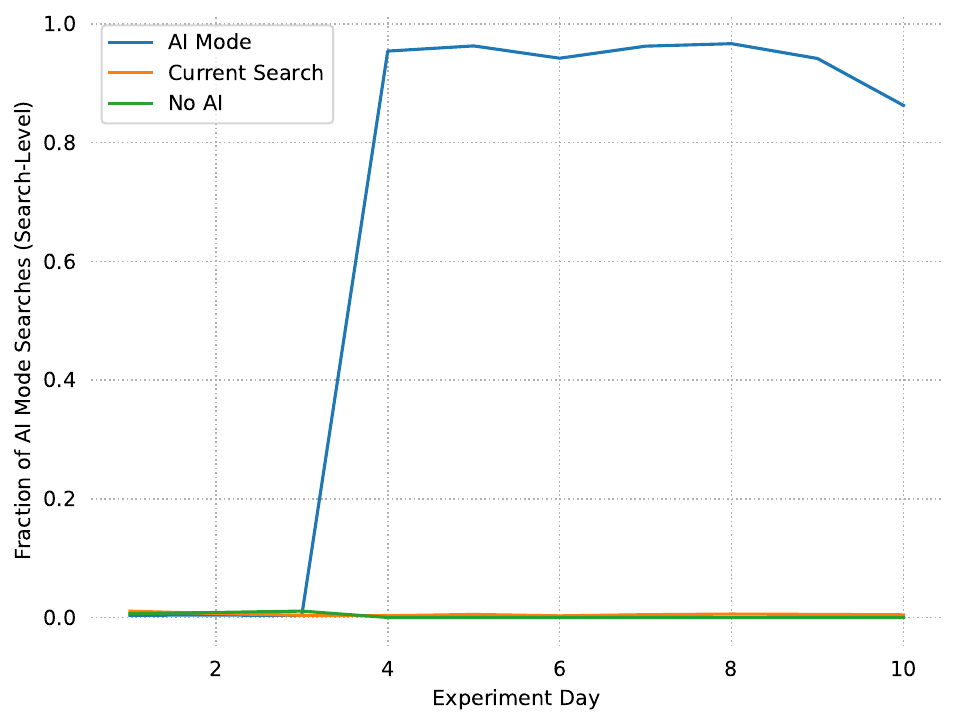}
    % Captions go below figures
    \caption{\textbf{Fraction of Searches Routed to AI Mode (Search-Level).} Each search is given equal weight.}
    \label{fig:ai_mode_compliance_search} % give each figure a logical label name
\end{figure}
Following our pre-registration, we measure the success of the No AI Search intervention by looking at the proportion of AI Overview HTML elements that were successfully hidden. We measure the success of the AI Mode intervention by looking at the proportion of Google searches routed to AI Mode according to URL parameters.
Figure \ref{fig:no_ai_compliance_time} shows the fraction of AI Overviews successfully hidden across experiment days (i.e., time relative to onboarding) in the No AI group. On the first experiment day, around 90\% of AI Overviews were successfully hidden. However, the success rate declined linearly as the experiment continued and was close to 0\% by the tenth experiment day. The reason for the declining success rate is that Google rolled out a change to the AI Overview HTML during the experiment. This broke our extension's logic for identifying and hiding AI Overviews. Figure \ref{fig:no_ai_compliance_users} shows the fraction of AI Overviews that were successfully hidden across participants. Overall, 51.1\% of AI Overviews were successfully hidden and the median participant had 50\% of AI Overviews hidden.
Figures \ref{fig:ai_mode_compliance_users} and \ref{fig:ai_mode_compliance_search} show the fraction of searches routed to AI Mode across treatment conditions at the user-level and search-level, respectively. During the baseline period, usage of AI Mode was extremely sparse across groups (0.6\% of searches). During the experiment, 94.7\% of searches in the AI Mode group were successfully routed to AI Mode. Six participants successfully accessed web search results (i.e., \url{google.com/search?q=testing&udm=14}) by using a non-standard URL structure (i.e., \url{google.com//search}) on which our intervention did not fire. As expected, 0.4\% of searches went to AI Mode in the Current Search group, where AI mode remained accessible, and 0\% of searches went to AI Mode in the No AI group.
Our pre-registration specified that we would report local average treatment effect (LATE) estimates if we detected non-compliance, where individual compliance was defined as receiving the assigned treatment on at least 90\% of searches. However, due to the breakdown of our No AI intervention, we instead measure compliance continuously, i.e., as the fraction of AI Overviews that were successfully hidden for each participant (see Figure \ref{fig:no_ai_compliance_users}). We measure compliance analogously in the AI Mode intervention, i.e., as the fraction of searches that were routed to AI Mode for each participant.
Although not pre-registered, we exclude users who did not run any searches that would have triggered an AI Overview from our LATE estimate comparing No AI Search to Current Search. Exposure is undefined for these users, and other than losing access to AI Mode, they could not have been affected by the No AI intervention. This subset constitutes 17.9\% of the No AI group and 16.3\% of the Current Search group. Using the same differential attrition tests that we use in the following section, we do not find evidence of differential rates ($p = 0.564$) or patterns ($p = 0.693$) among this subset.
\subsection{Differential attrition}
\label{si:attrition}
Following our pre-registration, we use randomization inference to test for differential attrition rates~\cite{lin2016green}. Specifically, we fit an OLS regression of attrition indicator on treatment assignment and calculate a heteroskedasticity-robust F-statistic for the hypothesis that all coefficients are equal to zero. We use randomization inference with 10,000 random assignments to calculate the p-value associated with the test statistic.
We also use randomization inference to test for differential attrition patterns~\cite{lin2016green}. Specifically, we fit an OLS regression of attrition indicator on treatment assignment, covariates, and treatment-covariate interactions and calculate a heteroskedasticity-robust F-statistic for the hypothesis that all interaction coefficients are equal to zero. We use the same set of covariates as in the covariate balance test. We use randomization inference with 10,000 random assignments to calculate the p-value associated with the test statistic.
Following our pre-registration, we measured attrition in two ways: (1) participants who did not complete the final survey and (2) participants with zero searches during the experiment period. The fraction of participants who did not complete the final survey was 32.5\% in the No AI group, 32.3\% in the Current Search group, and 36.6\% in the AI group. We do not find evidence of differential attrition rates ($p = 0.292$) or patterns ($p = 0.591$). The fraction of participants with zero searches during the experiment period was 24.0\% in the No AI group, 20.7\% in the Current Search group, and 26.5\% in the AI group. The difference between these rates is marginal ($p = 0.109$), but not significant at the 0.05 threshold \cite{lin2016green}, and we do not find evidence of differential attrition patterns ($p = 0.208$).
% \jgnote{Maybe: add Lee bounds for primary outcomes to hedge against concerns about marginally higher attrition in AI Mode group.}
\section{Supplementary Results}
\subsection{Robustness Checks}
\label{si:robustness}
\begin{figure}[htp] % Do not use \begin{figure*}
	\centering
    \includegraphics[width=\textwidth]{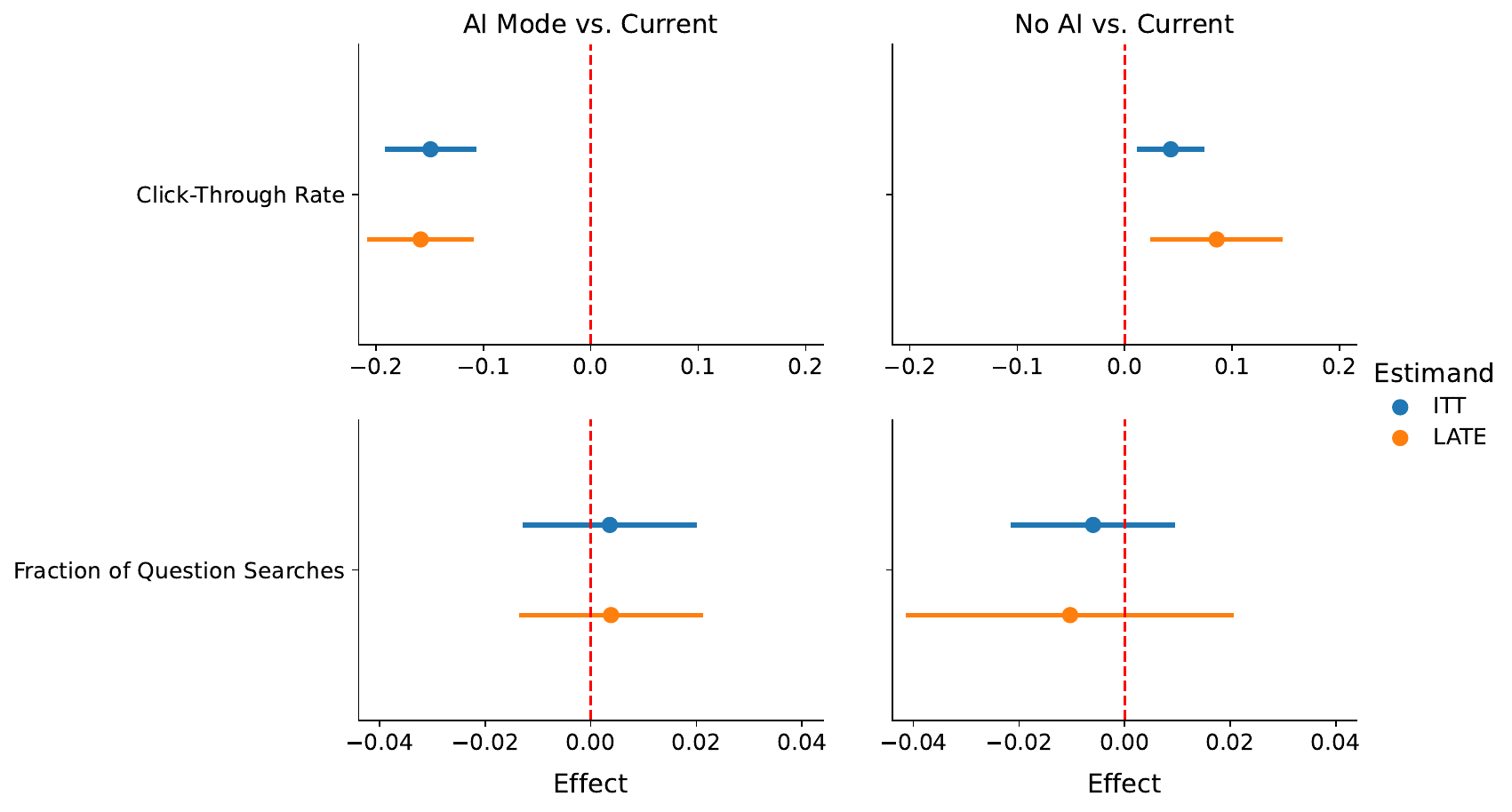}
    % Captions go below figures
    \caption{\textbf{Analysis of Click-Through Rate and Fraction of Question Searches using Search as the Analysis Unit.}}
    \label{fig:search_level} % give each figure a logical label name
\end{figure}
\begin{figure}[htp] % Do not use \begin{figure*}
	\centering
    \includegraphics[width=\textwidth]{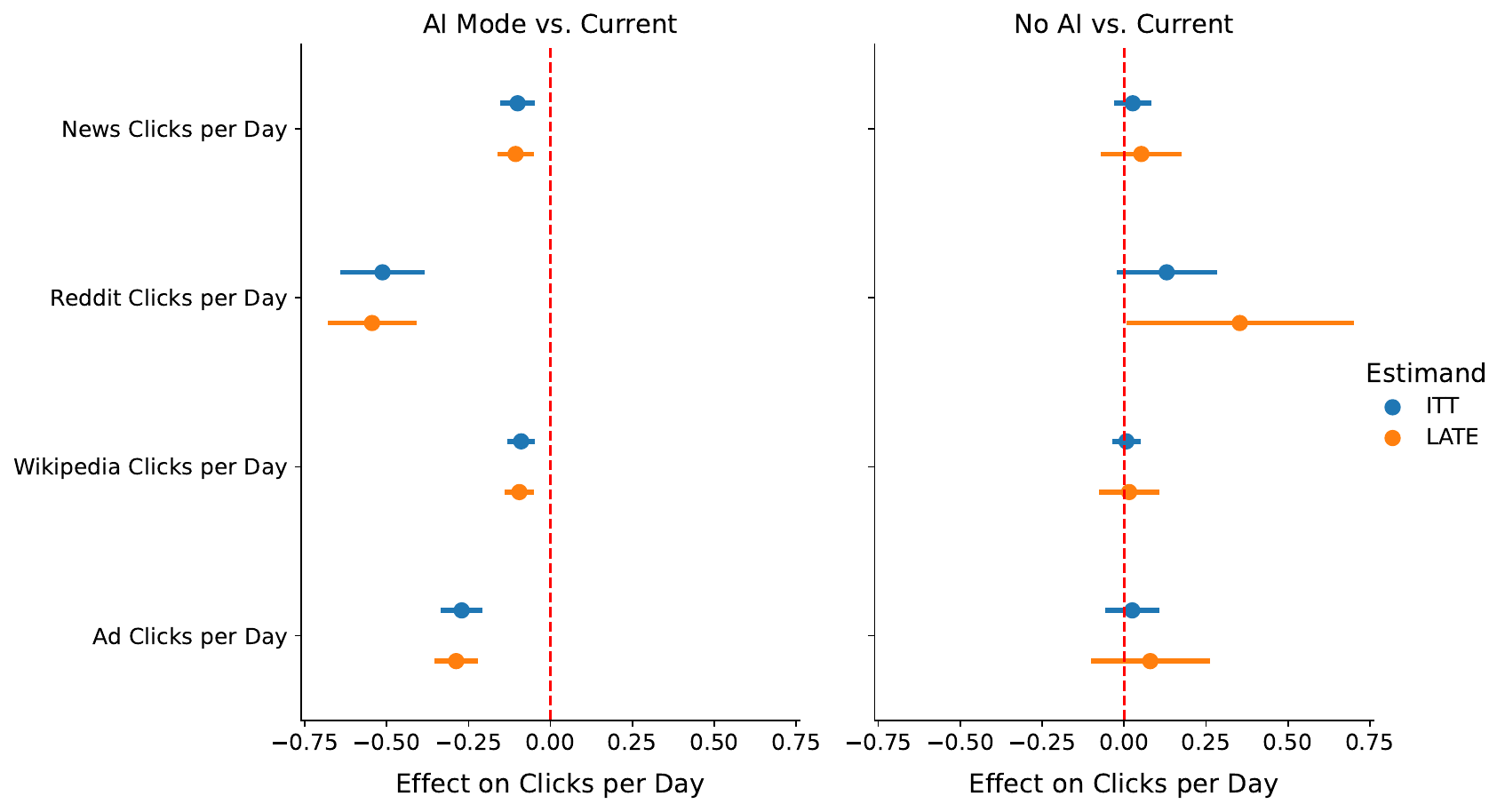}
    % Captions go below figures
    \caption{\textbf{Analysis of Secondary Click Outcomes using Clicks per Day.}}
    \label{fig:robustness_clicks} % give each figure a logical label name
\end{figure}
\begin{figure}[htp] % Do not use \begin{figure*}
	\centering
    \includegraphics[width=\textwidth]{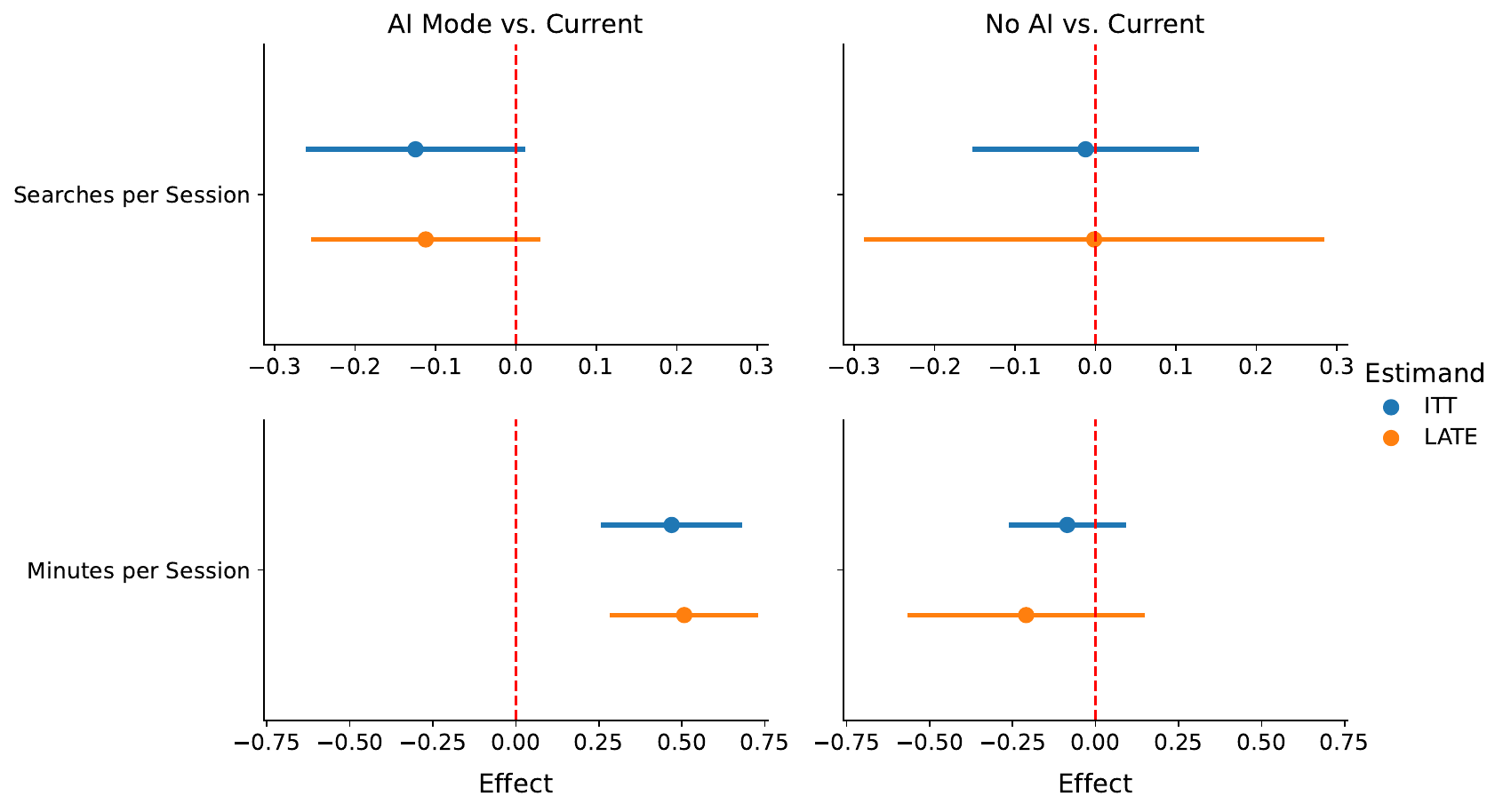}
    % Captions go below figures
    \caption{\textbf{Analysis of Searches per Session and Minutes per Session using Session as the Analysis Unit.}}
    \label{fig:session_level} % give each figure a logical label name
\end{figure}
\begin{figure}[htp] % Do not use \begin{figure*}
	\centering
    \includegraphics[width=\textwidth]{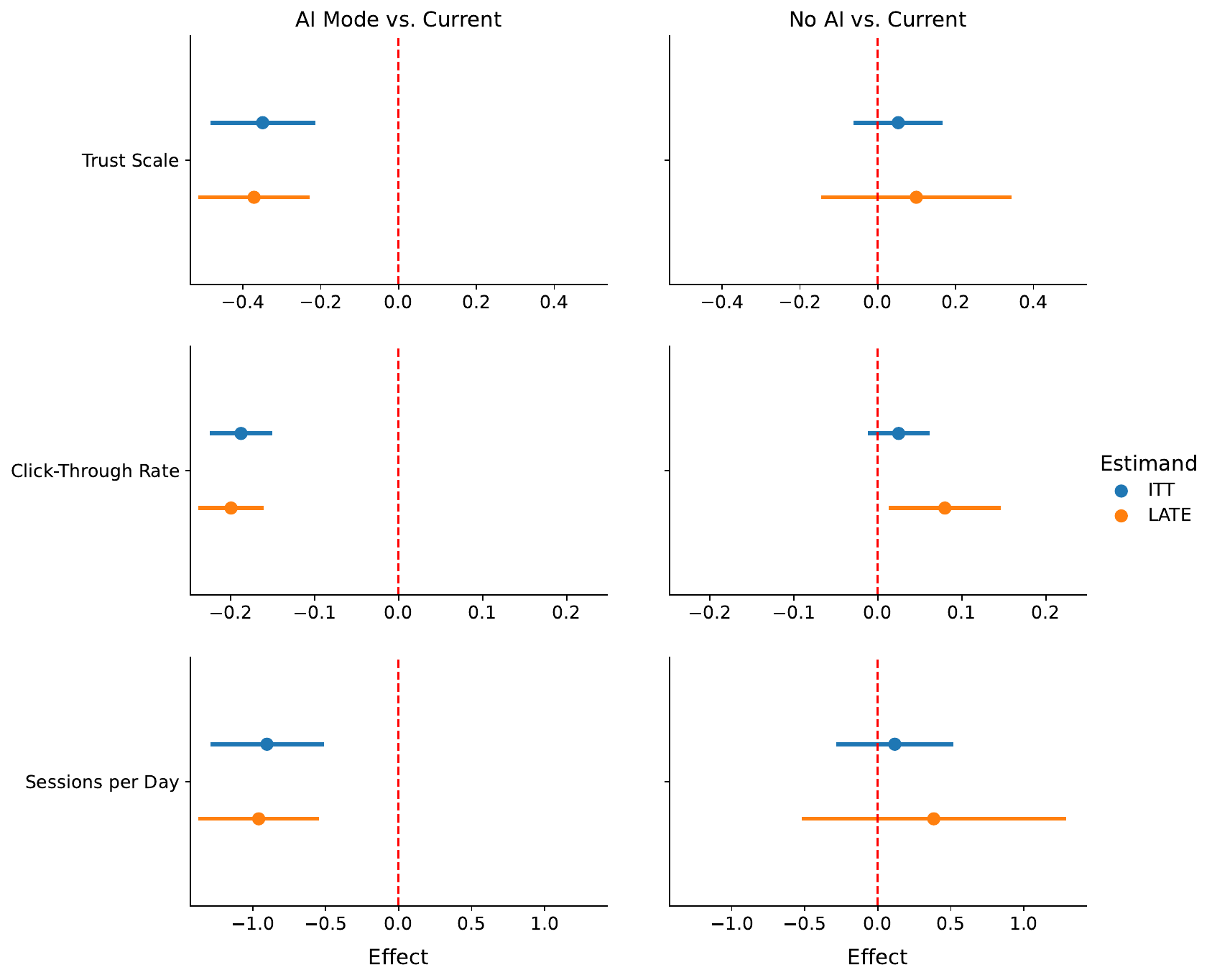}
    % Captions go below figures
    \caption{\textbf{Analysis of Primary Outcomes using Additional Controls Selected via Lasso.}}
    \label{fig:robustness_lasso} % give each figure a logical label name
\end{figure}
\begin{figure}[H] % Do not use \begin{figure*}
	\centering
    \includegraphics[width=\textwidth]{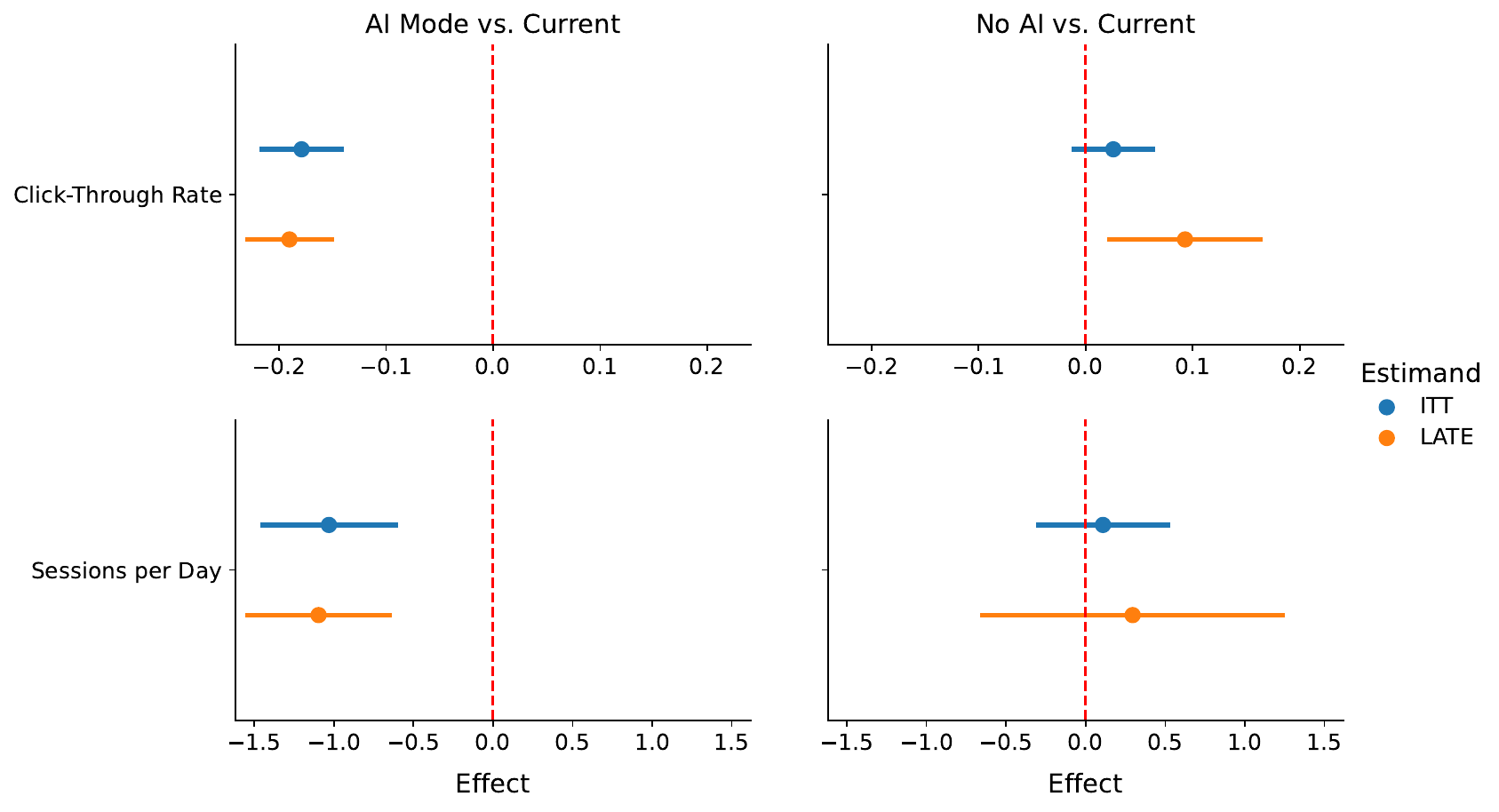}
    % Captions go below figures
    \caption{\textbf{Analysis of Primary Behavioral Outcomes among Subset that Completed the Final Survey (N = 956).}}
    \label{fig:robustness_sample} % give each figure a logical label name
\end{figure}
We conduct the following pre-registered robustness checks:
\begin{itemize}
    \item The effect of AI Mode on click-through rate is robust to counting each AI Mode conversation as one search, rather than counting each AI Mode prompt as one search (Table \ref{tab:robustness_search_def}).
    \item The effects of AI Mode and No AI on click-through rate are robust to analysis at the search-level instead of the user-level (Figure \ref{fig:search_level}, top panel). We analyzed click-through rate at the search-level by running OLS regression using the same controls and clustering standard errors at the user-level.
    \item The effects of AI Mode and No AI on sessions per day are robust to defining sessions using 3-minute boundaries and 10-minute boundaries instead of 5-minute boundaries (Table \ref{tab:robustness_session_def}).
    \item The effect of AI Mode on clicks to news sites is robust to using Google's Topics API (taxonomy 2, version 5) to classify news domains instead of \cite{yang2025newsdomains} (Table \ref{tab:robustness_news_def}).
    \item The effect of No AI on clicks to news sites is \textbf{not robust} to using Google's Topics API (taxonomy 2, version 5) to classify news domains instead of \cite{yang2025newsdomains} (Table \ref{tab:robustness_news_def}).
    % \jgnote{Comment on how these two news classifications differ.}
    \item The effects of AI Mode on clicks to news sites, Reddit, Wikipedia, and ads are robust to using clicks per day instead of fraction of users who clicked (Figure \ref{fig:robustness_clicks}, left panel).
    \item The effect of No AI on clicks to news sites is \textbf{not robust} to using clicks per day instead of fraction of users who clicked (Figure \ref{fig:robustness_clicks}, right panel). Effects on Wikipedia clicks per day and ad clicks per day are still null, while the effect on Reddit clicks per day is slightly positive.
    \item The effect of AI Mode on minutes per session is robust to analysis at the session-level instead of the user-level (Figure \ref{fig:session_level}, bottom-left panel). We analyzed minutes per session at the session-level by running OLS regression using the same controls and clustering standard errors at the user-level.
    \item The effect of No AI on minutes per session is \textbf{somewhat robust} to analysis at the session-level instead of the user-level (Figure \ref{fig:session_level}, bottom-right panel). Point estimates are still negative, but are no longer statistically significant.
    \item The effects of AI Mode and No AI on searches per session are also null when analyzed at the session-level instead of the user-level (Figure \ref{fig:session_level}, top panel).
    \item The effects of AI Mode and No AI on fraction of question searches are also null when analyzed at the search-level instead of the user-level (Figure \ref{fig:search_level}, bottom panel).
    \item The effect of AI Mode and No AI on primary outcomes is robust to using additional controls selected via Lasso regression (Figure \ref{fig:robustness_lasso}).
    \item The effect of AI Mode and No AI on primary behavioral outcomes is robust to restricting to the subset of participants ($N = 956$) that completed the final survey (Figure \ref{fig:robustness_sample}).
\end{itemize}
\begin{table}[ht] % Do not use \begin{table*}
	\centering
    \footnotesize
	% Captions go above tables
	\caption{\textbf{Click-Through Rate Estimates Using Different Definitions of AI Mode Searches.}}
	\label{tab:robustness_search_def}
    \scalebox{0.85}{
        \begin{tabular}{llllrrrlr}
\toprule
Outcome & Version & Comparison & Estimand & N & Estimate & SE & 95\% CI & p \\
\midrule
Click-Through Rate & 1 conversation = 1 search & AI Mode vs. Current & ITT & 714 & -0.149 & 0.018 & [-0.185, -0.113] & $<$0.001 \\
Click-Through Rate & 1 prompt = 1 search & AI Mode vs. Current & ITT & 714 & -0.188 & 0.018 & [-0.222, -0.153] & $<$0.001 \\
Click-Through Rate & 1 conversation = 1 search & AI Mode vs. Current & LATE & 714 & -0.158 & 0.019 & [-0.196, -0.120] & $<$0.001 \\
Click-Through Rate & 1 prompt = 1 search & AI Mode vs. Current & LATE & 714 & -0.199 & 0.019 & [-0.236, -0.163] & $<$0.001 \\
\bottomrule
\end{tabular}

    }
\end{table}
\begin{table}[ht] % Do not use \begin{table*}
	\centering
    \footnotesize
	% Captions go above tables
	\caption{\textbf{Sessions per Day Estimates Using Different Session Length Boundaries.}}
	\label{tab:robustness_session_def}
    \scalebox{0.85}{
        \begin{tabular}{llllrrrlr}
\toprule
Outcome & Version & Comparison & Estimand & N & Estimate & SE & 95\% CI & p \\
\midrule
Sessions per Day & 10 minutes & AI Mode vs. Current & ITT & 714 & -0.669 & 0.151 & [-0.964, -0.374] & $<$0.001 \\
Sessions per Day & 3 minutes & AI Mode vs. Current & ITT & 714 & -1.122 & 0.224 & [-1.562, -0.683] & $<$0.001 \\
Sessions per Day & 5 minutes & AI Mode vs. Current & ITT & 714 & -0.920 & 0.192 & [-1.296, -0.545] & $<$0.001 \\
Sessions per Day & 10 minutes & AI Mode vs. Current & LATE & 714 & -0.711 & 0.161 & [-1.026, -0.396] & $<$0.001 \\
Sessions per Day & 3 minutes & AI Mode vs. Current & LATE & 714 & -1.192 & 0.239 & [-1.661, -0.723] & $<$0.001 \\
Sessions per Day & 5 minutes & AI Mode vs. Current & LATE & 714 & -0.978 & 0.204 & [-1.379, -0.577] & $<$0.001 \\
Sessions per Day & 10 minutes & No AI vs. Current & ITT & 742 & 0.115 & 0.152 & [-0.183, 0.413] & 0.449 \\
Sessions per Day & 3 minutes & No AI vs. Current & ITT & 742 & 0.097 & 0.233 & [-0.360, 0.553] & 0.677 \\
Sessions per Day & 5 minutes & No AI vs. Current & ITT & 742 & 0.100 & 0.197 & [-0.287, 0.487] & 0.612 \\
Sessions per Day & 10 minutes & No AI vs. Current & LATE & 615 & 0.366 & 0.347 & [-0.316, 1.049] & 0.292 \\
Sessions per Day & 3 minutes & No AI vs. Current & LATE & 615 & 0.349 & 0.537 & [-0.705, 1.403] & 0.516 \\
Sessions per Day & 5 minutes & No AI vs. Current & LATE & 615 & 0.345 & 0.453 & [-0.545, 1.236] & 0.446 \\
\bottomrule
\end{tabular}

    }
\end{table}
\begin{table}[ht] % Do not use \begin{table*}
	\centering
    \footnotesize
	% Captions go above tables
	\caption{\textbf{Fraction of Users Clicking to News Using Different News Domain Labels.}}
	\label{tab:robustness_news_def}
    \scalebox{0.85}{
        \begin{tabular}{llllrrrlr}
\toprule
Outcome & Version & Comparison & Estimand & N & Estimate & SE & 95\% CI & p \\
\midrule
Fraction of News Clickers & Topics API & AI Mode vs. Current & ITT & 714 & -0.081 & 0.017 & [-0.114, -0.048] & $<$0.001 \\
Fraction of News Clickers & Yang (2026) & AI Mode vs. Current & ITT & 714 & -0.125 & 0.032 & [-0.187, -0.063] & $<$0.001 \\
Fraction of News Clickers & Topics API & AI Mode vs. Current & LATE & 714 & -0.086 & 0.018 & [-0.121, -0.051] & $<$0.001 \\
Fraction of News Clickers & Yang (2026) & AI Mode vs. Current & LATE & 714 & -0.133 & 0.034 & [-0.199, -0.067] & $<$0.001 \\
Fraction of News Clickers & Topics API & No AI vs. Current & ITT & 742 & 0.001 & 0.019 & [-0.036, 0.037] & 0.967 \\
Fraction of News Clickers & Yang (2026) & No AI vs. Current & ITT & 742 & 0.071 & 0.033 & [0.006, 0.137] & 0.033 \\
Fraction of News Clickers & Topics API & No AI vs. Current & LATE & 615 & 0.003 & 0.043 & [-0.082, 0.088] & 0.940 \\
Fraction of News Clickers & Yang (2026) & No AI vs. Current & LATE & 615 & 0.202 & 0.073 & [0.058, 0.346] & 0.006 \\
\bottomrule
\end{tabular}

    }
\end{table}
\begin{figure}[ht] % Do not use \begin{figure*}
	\centering
    \includegraphics[width=\textwidth]{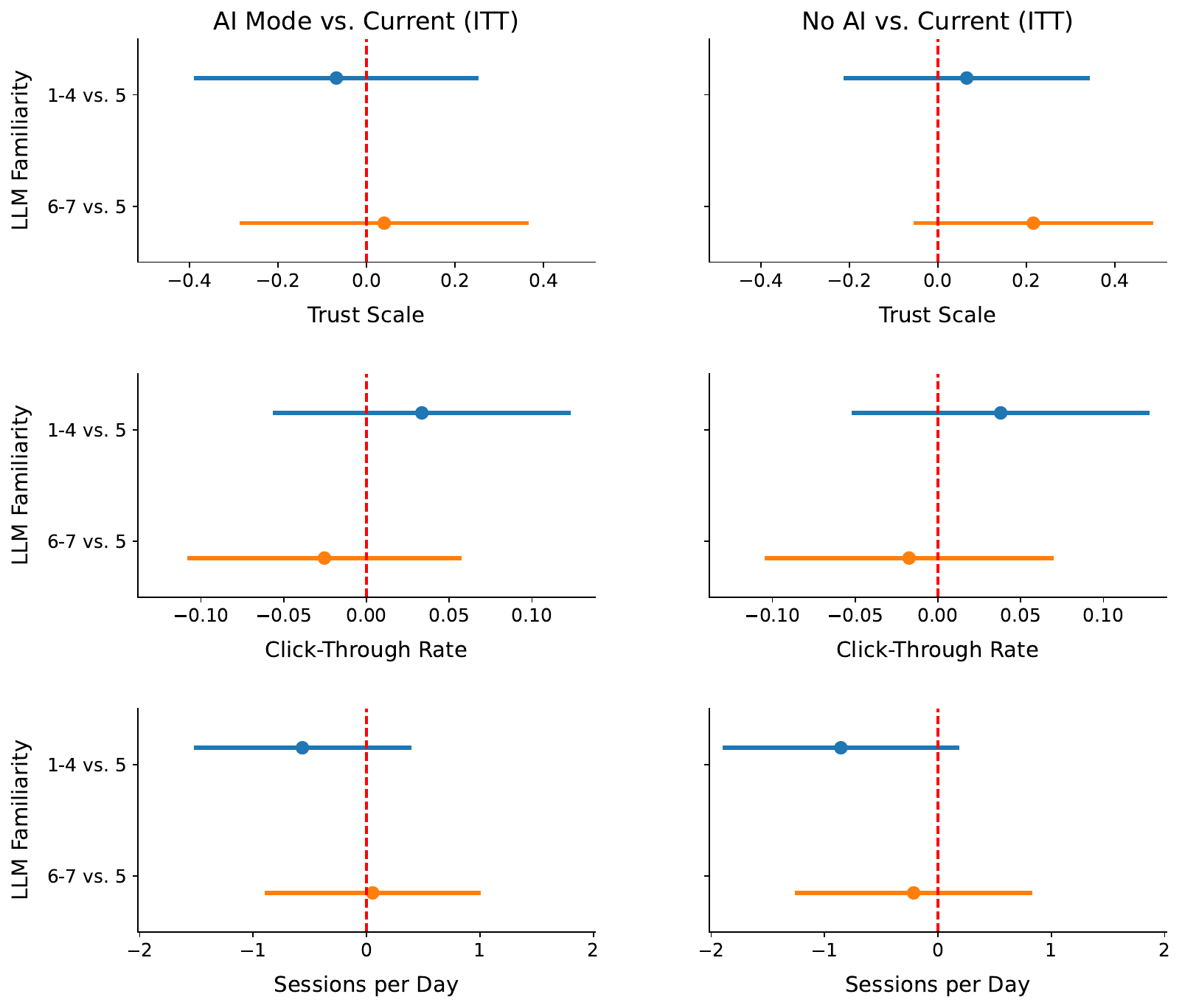}
    % Captions go below figures
    \caption{\textbf{Heterogeneous Effects of AI Mode Search and No AI Search on Primary Outcomes Across Baseline Levels of LLM Familiarity.}}
    \label{fig:hte_llm_familiarity} % give each figure a logical label name
\end{figure}
\begin{figure}[ht] % Do not use \begin{figure*}
	\centering
    \includegraphics[width=\textwidth]{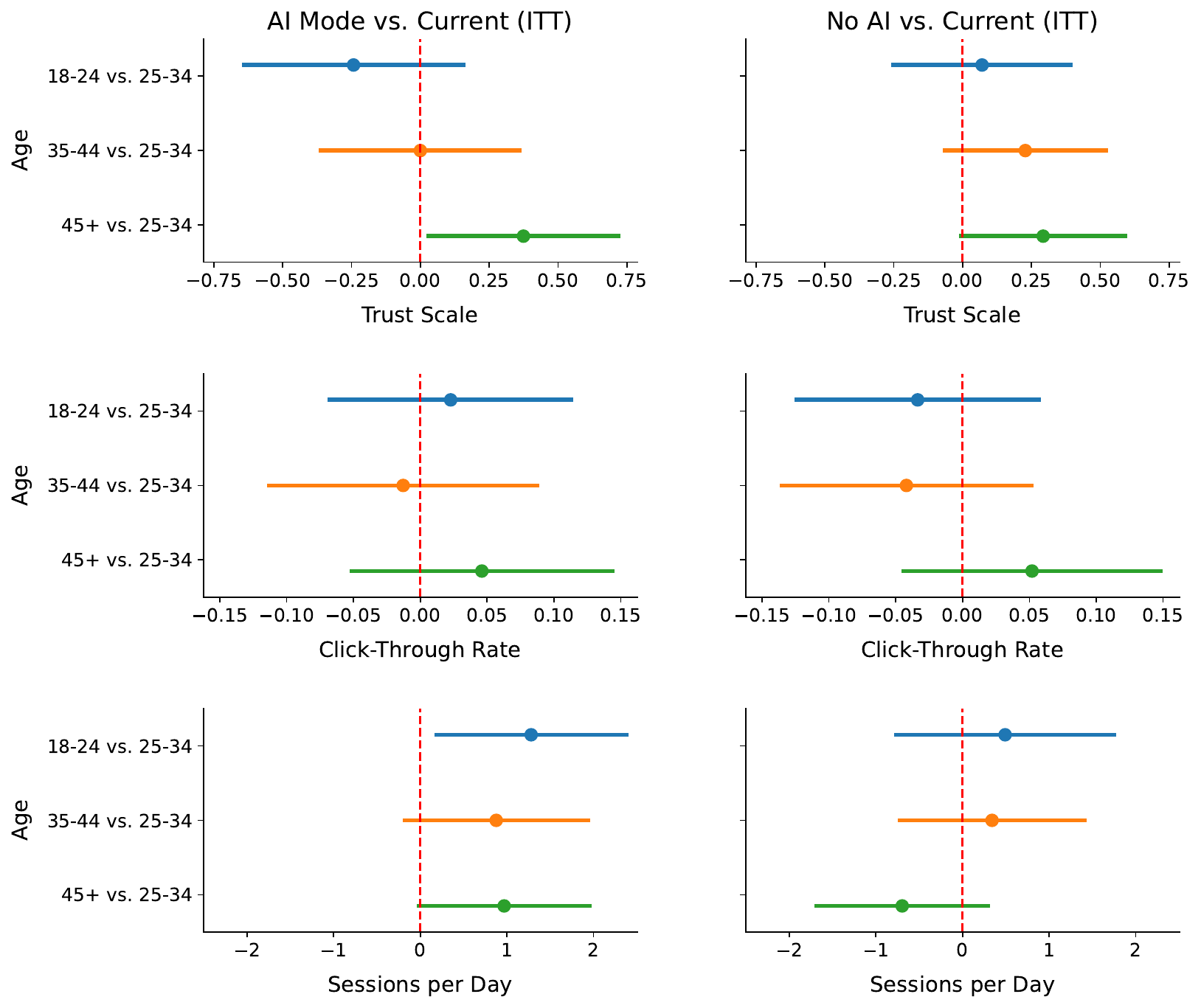}
    % Captions go below figures
    \caption{\textbf{Heterogeneous Effects of AI Mode Search and No AI Search on Primary Outcomes Across Age Groups.}}
    \label{fig:hte_age} % give each figure a logical label name
\end{figure}
\begin{figure}[ht] % Do not use \begin{figure*}
	\centering
    \includegraphics[width=\textwidth]{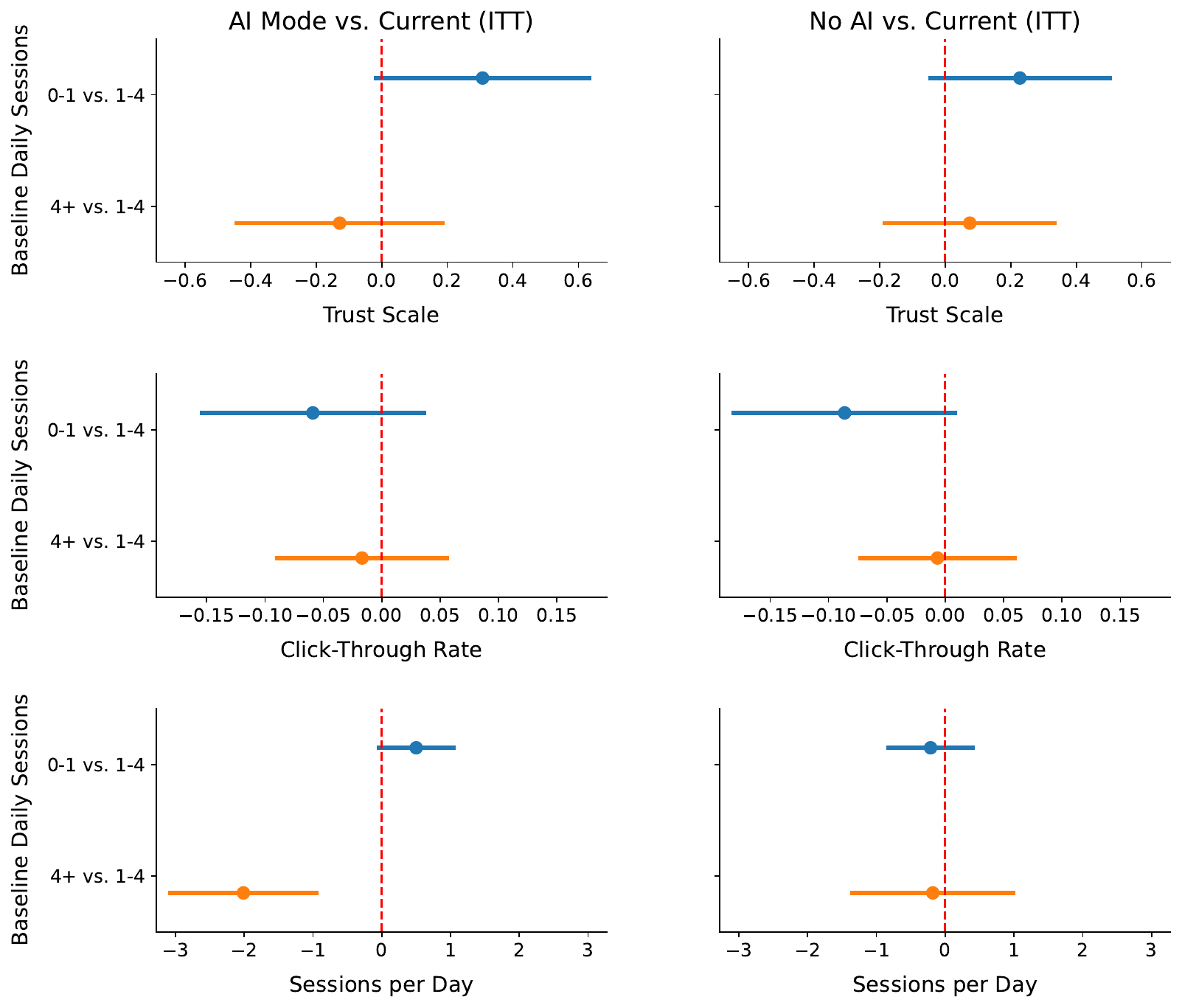}
    % Captions go below figures
    \caption{\textbf{Heterogeneous Effects of AI Mode Search and No AI Search on Primary Outcomes Across Baseline Number of Daily Search Sessions.}}
    \label{fig:hte_n_sessions_per_day_pre} % give each figure a logical label name
\end{figure}
\subsection{Heterogeneous Treatment Effects}
\label{si:hte}
We test for heterogeneous effects on our primary outcomes across two pre-registered moderators: (1) pre-treatment familiarity with LLMs (coarsened) and (2) pre-treatment number of Google sessions (coarsened). We also test for heterogeneous effects across one exploratory moderator: age.
We analyze heterogeneous effects by running OLS regression of the outcome on participants’ treatment assignment, moderator, and treatment-moderator interactions. We run one regression per moderator and report treatment-moderator coefficients. We control for the baseline value of the outcome and recruitment channel, and use HC2 standard errors.
We coarsen LLM familiarity into three groups: not familiar at all, slightly familiar, somewhat familiar, and moderately familiar ($N = 343$), very familiar ($N = 312$), highly familiar and extremely familiar ($N = 445$). We coarsen pre-treatment daily search sessions into three groups: up to one daily session ($N = 366$), 1-4 daily sessions ($N = 374$), more than 4 daily sessions ($N = 360$). We coarsen age into four groups: 18--24 ($N = 233$), 25--34 ($N = 340$), 35-44 ($N = 254$), and 45+ ($N = 273$). Figure \ref{fig:hte_llm_familiarity} shows the results for LLM familiarity: no interactions are significant. Figure \ref{fig:hte_n_sessions_per_day_pre} shows the results for baseline daily sessions. AI Mode reduces the number of sessions per day significantly more for heavy Google users (i.e., more than 4 daily sessions) (estimate = -2.01 sessions per day, P = 0.005). All other interactions are insignificant after adjusting for multiple comparisons. Figure \ref{fig:hte_age} shows the results for age: no interactions are significant after adjusting for multiple comparisons.
\clearpage
\section{Survey Questionnaires}
% ============================================================
% Survey questions for search_experiment
% Sources: onboard/forms.py (InformedConsentForm, DemographicsForm),
%          frontend/forms.py (IntroSurveyForm, FinalSurveyForm)
% Active (non-commented-out) fields only.
% ============================================================
\subsection{Onboarding}
\subsubsection*{Informed Consent}
\begin{enumerate}
    \item I am living in the United States. \hfill \textit{(checkbox)}
    \item I am 18 years old or older. \hfill \textit{(checkbox)}
    \item My main web browser is Google Chrome. \hfill \textit{(checkbox)}
    \item My main search engine is Google Search. \hfill \textit{(checkbox)}
    \item I have read the above information and consent to participate. \hfill \textit{(checkbox)}
    \item Please enter your 24 character Prolific ID. \hfill \textit{(free text)}
\end{enumerate}
\subsubsection*{Demographics}
\label{si:demographics-survey}
\begin{enumerate}
    \item What's your age?
    \begin{itemize}
        \item 18 -- 24
        \item 25 -- 34
        \item 35 -- 44
        \item 45 -- 54
        \item 55 -- 64
        \item 65 -- 74
        \item 75 -- 84
        \item 85 or older
    \end{itemize}
    \item What's your gender identity?
    \begin{itemize}
        \item Woman
        \item Man
        \item Non-binary/third gender
        \item Prefer not to say
    \end{itemize}
    \item What's your race and/or ethnicity? Select all that apply:
    \begin{itemize}
        \item American Indian or Alaska Native
        \item Black or African American
        \item Hispanic or Latino
        \item Indian Subcontinent
        \item Middle Eastern or North African
        \item Native Hawaiian or Pacific Islander
        \item Southeast Asian or East Asian
        \item White
        \item Other/Not listed
    \end{itemize}
    \item What's your level of education?
    \begin{itemize}
        \item Less than high school
        \item High school graduate
        \item Some college
        \item Bachelor's degree
        \item Master's degree or higher
    \end{itemize}
    \item What's your annual household income before taxes?
    \begin{itemize}
        \item Less than \$25,000
        \item \$25,000 -- \$49,999
        \item \$50,000 -- \$99,999
        \item \$100,000 -- \$199,999
        \item \$200,000 or more
    \end{itemize}
    \item What's your political party identification?
    \begin{itemize}
        \item Strong Democrat
        \item Not very strong Democrat
        \item Lean Democrat
        \item Independent
        \item Lean Republican
        \item Not very strong Republican
        \item Strong Republican
    \end{itemize}
    \item What is your 5 digit zip code? \hfill \textit{(free text)}
\end{enumerate}
\subsection{Pre-Experiment Survey}
\label{si:pre-survey}
\begin{enumerate}
    \item How familiar are you with LLM applications such as ChatGPT, Claude, and Gemini?
    \begin{itemize}
        \item Not familiar at all
        \item Slightly familiar
        \item Somewhat familiar
        \item Moderately familiar
        \item Very familiar
        \item Highly familiar
        \item Extremely familiar
    \end{itemize}
    \item How often do you use LLM applications such as ChatGPT, Claude, and Gemini?
    \begin{itemize}
        \item Never
        \item At least once per month
        \item At least once per week
        \item At least once per day
        \item Multiple times per day
    \end{itemize}
    \item Overall, how do you feel about LLM applications such as ChatGPT, Claude, and Gemini?
    \begin{itemize}
        \item Negative
        \item Somewhat negative
        \item Neutral
        \item Somewhat positive
        \item Positive
    \end{itemize}
    \item When you search on Google, how believable is the information you find?
    \begin{itemize}
        \item Not at all
        \item Slightly
        \item Somewhat
        \item Moderately
        \item Very
        \item Highly
        \item Extremely
    \end{itemize}
    \item When you search on Google, how trustworthy is the information you find? \hfill \textit{(same 7-point scale as above)}
    \item When you search on Google, how accurate is the information you find? \hfill \textit{(same 7-point scale as above)}
    \item When you search on Google, how biased is the information you find? \hfill \textit{(same 7-point scale as above)}
    \item When you search on Google, how complete is the information you find? \hfill \textit{(same 7-point scale as above)}
    \item Please select `Slightly' for this question. \hfill \textit{(attention check; same 7-point scale as above)}
    \item I have agency over the information-seeking process on Google.
    \begin{itemize}
        \item Strongly disagree
        \item Disagree
        \item Somewhat disagree
        \item Neither agree nor disagree
        \item Somewhat agree
        \item Agree
        \item Strongly agree
    \end{itemize}
    \item Information-seeking on Google leads to results that are personalized and relevant to me. \hfill \textit{(same 7-point agreement scale as above)}
    \item In general, information-seeking on Google is: \hfill \textit{(9 semantic differential items, 5-point scale, no intermediate labels)}
    \begin{itemize}
        \item Useless -- Useful
        \item Unpleasant -- Pleasant
        \item Bad -- Good
        \item Annoying -- Nice
        \item Ineffective -- Effective
        \item Irritating -- Likeable
        \item Worthless -- Helpful
        \item Undesirable -- Desirable
        \item Sleep-inducing -- Stimulating
    \end{itemize}
\end{enumerate}
\subsection{Post-Experiment Survey}
\label{si:post-survey}
\begin{enumerate}
    \item \textit{(For each head-to-head comparison sample, response order randomized as Response A/B)}
    \begin{enumerate}
        \item Which of the two responses do you prefer?
        \item Which of the two responses do you find more trustworthy?
        \item Why did you find Response A more trustworthy? \hfill \textit{(free text)}
    \end{enumerate}
    \item What was your overall impression of Google's AI Mode after one week of using it? \hfill \textit{(AI Mode participants only; free text)}
    \item Over the last 7 days, how believable was the information you found searching on Google?
    \begin{itemize}
        \item Not at all
        \item Slightly
        \item Somewhat
        \item Moderately
        \item Very
        \item Highly
        \item Extremely
    \end{itemize}
    \item Over the last 7 days, how trustworthy was the information you found searching on Google? \hfill \textit{(same 7-point scale as above)}
    \item Over the last 7 days, how accurate was the information you found searching on Google? \hfill \textit{(same 7-point scale as above)}
    \item Over the last 7 days, how biased was the information you found searching on Google? \hfill \textit{(same 7-point scale as above)}
    \item Over the last 7 days, how complete was the information you found searching on Google? \hfill \textit{(same 7-point scale as above)}
    \item Over the last 7 days, I had agency over the information-seeking process on Google.
    \begin{itemize}
        \item Strongly disagree
        \item Disagree
        \item Somewhat disagree
        \item Neither agree nor disagree
        \item Somewhat agree
        \item Agree
        \item Strongly agree
    \end{itemize}
    \item Over the last 7 days, information-seeking on Google led to results that were personalized and relevant to me. \hfill \textit{(same 7-point agreement scale as above)}
    \item In general, information-seeking on Google is: \hfill \textit{(9 semantic differential items, 5-point scale, no intermediate labels)}
    \begin{itemize}
        \item Useless -- Useful
        \item Unpleasant -- Pleasant
        \item Bad -- Good
        \item Annoying -- Nice
        \item Ineffective -- Effective
        \item Irritating -- Likeable
        \item Worthless -- Helpful
        \item Undesirable -- Desirable
        \item Sleep-inducing -- Stimulating
    \end{itemize}
    \item If your search experience from the last 7 days remained permanent, how likely would you be to switch to Bing?
    \begin{itemize}
        \item Extremely unlikely
        \item Unlikely
        \item Somewhat unlikely
        \item Neither likely nor unlikely
        \item Somewhat likely
        \item Likely
        \item Extremely likely
    \end{itemize}
    \item Did you search on a browser other than this one in the past week?
    \begin{itemize}
        \item Yes
        \item No
    \end{itemize}
\end{enumerate}
% \subsubsection*{Descriptive Results}
% \jgnote{Possible things to include}:
% \begin{itemize}
%     \item Baseline clicks across parts of SERP
%     \item Baseline clicks across domains
%     \item AI Overview effects split at page-level
%     \item Impacts on fraction of navigational searches
%     \item \swnote{Impacts on search query length}~\cite{reid_ai_2025}
%     \item Impacts on LLM conversations
%     \item Survey: whether people used another browser
%     \item Behavioral: fraction of people who searched for "ai mode"
% \end{itemize}
%%%%%%%%%%%%%%%% SUPPLEMENTARY TEXT %%%%%%%%%%%%%%%
% \subsection*{Supplementary Text}
% The Supplementary Text section can only be used to directly support statements made in the main text
% e.g. to present more detailed justifications of assumptions, investigate alternative scenarios,
% provide extended acknowledgements etc.
% Material in this section cannot claim results or conclusions that weren't mentioned in the main text.
% To refer to this section from the main text, just write (Supplementary Text).
% \subsubsection*{Example supplement heading}
% The two main sections of the supplement can be split up using headings.
% If your supplement is very short you might need to uncomment the following line to avoid
% layout problems with the figures and tables.
\newpage

\end{document}